\documentclass[journal,10pt]{IEEEtran}
\usepackage{amsfonts}
\IEEEoverridecommandlockouts

\ifCLASSINFOpdf
\else
\fi
\usepackage{epsfig}
\usepackage{graphicx}
\usepackage{subfigure}
\usepackage{psfig}
\usepackage{epsf}
\usepackage[cmex10]{amsmath}
\usepackage{booktabs}
\usepackage{fancyhdr}
\usepackage{stfloats}
\usepackage{color}
\usepackage{amssymb}
\usepackage{setspace}
\usepackage{bm}

\begin{document}
 \title{Robust Trajectory and Transmit Power Optimization for Secure UAV-Enabled Cognitive Radio Networks}
\author{\IEEEauthorblockN{Yifan Zhou, Fuhui Zhou, \IEEEmembership{Member,~IEEE,} Huilin Zhou, Derrick Wing Kwan Ng, \IEEEmembership{Senior Member, IEEE,} \\ and Rose Qingyang Hu, \IEEEmembership{Fellow, IEEE}}

\thanks{Manuscript received Sep. 29, 2019; revised Jan. 28, 2020 and accepted Mar. 02, 2020. Date of publication ****; date of current version ****. The editor coordinating the review of this paper and approving it for publication was Prof. MD Jahangir Hossain. }

\thanks{The research was supported in part by the Natural Science Foundation of China (61561034, 61701214, 61631020, 61827801, 61931011, and 61901216), in part by the National Key Research and Development Project of China under Grant 2018YFB1800801, the National Science Foundation under Grants EECS-1308006, the Young Natural Science Foundation of Jiangxi Province  under Grant 2015BAB207001, in part by the Postgraduate Innovation Special Foundation of Nanchang University under Grant CX2018157, in part by the US National Science Foundation under Grant EARS1547312, in part by the Excellent Youth Foundation of Jiangxi Province under Grant 2018ACB21012 and in part by Young Elite Scientist Sponsorship Program by CAST. \emph{Corresponding author:} Fuhui Zhou and Huilin Zhou.)

Y. Zhou and H. Zhou are with the School of Information Engineering, Nanchang University, Nanchang 330031, China (e-mail: zhouyifan@email.ncu.edu.cn; zhouhuilin@ncu.edu.cn).

Fuhui Zhou is with the College of Electronic and Information Engineering, Nanjing University of Aeronautics and Astronautics, Nanjing, 210000, P. R. China. He is also with Key Laboratory of Dynamic Cognitive System of Electromagnetic Spectrum Space (Nanjing University of Aeronautics and Astronautics) and with Ministry of Industry and Information Technology, Nanjing 211106, China (e-mail: zhoufuhui@ieee.org).

R. Q. Hu is with the Department of Electrical and Computer Engineering, Utah State University, Logan, UT 84322 USA (e-mail: rosehu@ieee.org).

D. W. K. Ng is with the University of New South Wales, Sydney, NSW 2052, Australia (e-mail: w.k.ng@unsw.edu.au).
}}

\maketitle
\begin{abstract}
Cognitive radio is a promising  technology to improve spectral efficiency. However, the secure performance of a secondary network achieved by using physical layer security techniques is limited by its transmit power and channel fading. In order to tackle this issue, a cognitive unmanned aerial vehicle (UAV) communication network is studied by exploiting the high flexibility of a UAV and the possibility of establishing line-of-sight links. The average secrecy rate of the secondary network is maximized by robustly optimizing the UAV's trajectory and transmit power. Our problem formulation takes into account two practical inaccurate location estimation cases, namely, the worst case and the outage-constrained case. In order to solve those challenging non-convex problems, an iterative algorithm based on $\mathcal{S}$-Procedure is proposed for the worst case while an iterative algorithm based on Bernstein-type inequalities is proposed for the outage-constrained case. The proposed algorithms can obtain effective suboptimal solutions of the corresponding problems. Our simulation results demonstrate that the algorithm under the outage-constrained case can achieve a higher average secrecy rate with a low computational complexity compared to that of the algorithm under the worst case. Moreover, the proposed schemes can improve the secure communication performance significantly compared to other benchmark schemes.
\end{abstract}
\IEEEpeerreviewmaketitle

\section{Introduction}

\subsection{Background and Motivation}
The explosive increase of wide-band service requirements and the unprecedented proliferation of mobile devices result in a severe spectrum scarcity issue. In order to alleviate the spectrum crunch and to meet the increasing demand for high data rates, cognitive radio (CR) that aims to realize spectrum sharing between a primary network and a secondary network has been proposed to improve the spectral efficiency \cite{S. Haykin}, \cite{A. Goldsmith}. Specifically, secondary users (SUs) coexist with primary users (PUs) in the same frequency spectrum while ensuring that PUs can tolerate the interference caused by SUs. Due to the promised high-spectral efficiency brought by CR, it has been widely investigated in traditional terrestrial communication networks such as cellular networks, wireless sensor networks, and relaying networks \cite{O. B. Akan}, \cite{X. Huang}. However, in CR networks, the transmit power of a secondary base station should be restricted in order to guarantee the quality of service (QoS) of the PUs. What's worse, the harsh channel fading conditions of terrestrial communication networks further degrade the performance of SUs. Thus, it is of utmost importance to study how to improve the performance of SUs in CR networks.

Recently, unmanned aerial vehicles (UAVs) communications have been extensively investigated for wireless communications in both military and civil applications due to the advantages of UAVs in terms of highly controllable mobility and flight flexibility \cite{L. Gupta}-\cite{Y. Sun}. Compared with terrestrial CR networks, which are dominated by non-line-of-sight (NLoS) links due to multipaths and blockages, UAV-enabled CR networks can flexibly deploy UAVs in the air according to the actual environment \cite{X. Lin}. In particular, the communication performance of ground nodes can be greatly improved since the air-to-ground channel is dominated by line-of-sight (LoS) with generally a small path loss exponent \cite{A. A. Khuwaja}. Moreover, the LoS channel facilitates the establishment of high data rate communications between UAVs and ground nodes. In practice, the throughput of UAV-enabled CR networks can be significantly improved by designing the UAV's trajectory that exploits the extra design degrees of freedom \cite{S. Hayat}, \cite{Q. Wu}. Thus, it is promising to exploit UAVs to establish CR networks that can improve the performance of SUs. However, due to the existence of strong LoS links from UAVs to PUs in UAV-enabled CR networks, the interference caused by the cognitive UAV transmitters to PUs is also increased substantially \cite{Y. Huang}. Thus, this is a major challenge for the design of the UAV's trajectory and transmission power in UAV-enabled CR networks.

Moreover, due to the broadcast nature of wireless communications and the existence of LoS channels, potential eavesdroppers (Eves) on the ground are more prone to illegally intercept confidential information sent to legitimate recipients. It results in a significant security challenge for UAV-enabled CR networks. Besides conventional cryptography encryption methods enabling secure information transmission, physical-layer security technology has emerged as a promising alternative approach to guarantee the security of wireless communication systems by exploiting the physical layer characteristics of wireless channels \cite{P. K. Gopala}-\cite{X. Sun}. In \cite{Xiaobo Zhou}, the physical layer security of a UAV network was studied to jointly design the trajectory and transmission power of the UAV in order to maximize the minimum average secrecy rate of all users. As a significant metric for physical layer security, it was shown that the secrecy rate is closely related to channel conditions \cite{Q. Li}. Thus, it is necessary to design the trajectory in UAV-enabled CR networks to improve the secure performance of SUs.

To unlock the potential of UAV-enabled communication systems, joint UAV's trajectory and transmit power design has been studied under different scenarios \cite{C. Zhang}-\cite{A. M. Almasoud}. However, these works assumed perfect channel estimations, which depends on the availability of accurate estimation for the locations of ground nodes. In practice, it is extremely difficult to acquire the exact location of PUs/Eves due to the existence of location estimation errors and quantization errors. Especially in UAV-enabled CR networks, it is challenging for a UAV to keep tracking the exact location of PUs/Eves via a camera or a synthetic aperture radar due to the energy limitation of the UAV \cite{C. J. Li}. As a result, the trajectory design and power allocation based on the assumption of perfect location information may lead to a large performance loss for UAV-enabled CR networks in practice. Recently, although the robust trajectory and transmit power designs have been considered in \cite{M. Cui}, \cite{X. Sun}, only the worst-case scenario has been studied. Specifically, in the worst-case scenario, the location errors are modeled by bounded sets. However, the bounded location error model usually leads to a conservative resource allocation which underestimates the actual communication performance of UAV systems. Thus, in the considered UAV-enabled CR network, we not only focus on the robust design based on the bounded location error model, but also consider the robust design based on the probabilistic location error model. For the latter model, the location errors are captured by a probability distribution \cite{X. Zhou}, and outage probability constraints are proposed to replace the excessively conservative worst-case constraints adopted in the existing robust designs. Under this model, UAV-enabled CR networks can achieve better performance. In particular, the latter robust design is more suitable for scenarios with delay-sensitive communication devices \cite{F. Zhou}. Our proposed two robust UAV's trajectory and transmit power optimization schemes in this paper are promising to improve the secure communication performance of SUs against the channel estimation error and offer a better protection to the performance of PUs. Moreover, the studied fair performance comparison between the two proposed robust designs is helpful for the performance analysis in UAV-enabled CR networks with practically imperfect channel state information (CSI).

\begin{spacing}{-0.5}
\end{spacing}

\subsection{Related Works}

\subsubsection{Traditionally terrestrial CR networks}
The designs of robust resource allocation in traditionally terrestrial CR communication networks have been investigated under the bounded CSI error model in \cite{D. W. K. Ng}, \cite{Yanan Wu} and under the probabilistic CSI error model in \cite{S. Ma}-\cite{H.-W. Lee}. Specifically, in \cite{D. W. K. Ng}, the worst-case robust secure multiobjective resource allocation scheme was studied for multiple-input single-output CR networks with simultaneously wireless information and power transfer. Beamforming was designed to strike the tradeoff between the considered conflicting system design objectives. In \cite{Yanan Wu}, the authors studied the worst-case robust design of secure wireless information and power transfer in a cognitive relaying system. It was shown that by introducing a joint robust beamforming, transformation matrix and power splitting method, the minimization of transmit power can be achieved in the worst-case scenario. In order to improve the performance of CR networks under the bounded CSI error model, robust resource allocation designs under the outage probability constraints were proposed \cite{S. Ma}-\cite{H.-W. Lee}. In \cite{S. Ma}, a robust beamforming problem was studied in CR networks to minimize the total SU's transmit power via considering probabilistic chance constraints. It was shown that the robust beamforming design taking into account chance constraints is an effective approach to improve the performance compared to the conventional worst-case based method. Under the PU's performance outage probability constraints, a robust cooperative beamforming was optimized for a CR relaying network in \cite{S. Singh}. In \cite{H.-W. Lee}, the SU power allocation problem was studied in a CR network with uncertain knowledge of interference information. In \cite{F. Zhou} and \cite{H. Sun}, robust beamforming designs were considered under two different CSI error models in wireless power transfer systems. However, the robust resource allocation schemes proposed in \cite{D. W. K. Ng}-\cite{H. Sun} are not applicable to UAV-enabled CR networks, since the channels in UAV-enabled CR networks are LoS channels and are coupled with the trajectory of UAVs.

\subsubsection{UAV-enabled CR networks}
To improve the throughput of SUs in CR networks, UAV-enabled CR communication systems were studied in \cite{C. Zhang}-\cite{A. M. Almasoud}. In \cite{C. Zhang}, the authors investigated spectrum sharing in UAV-enabled small-cell networks to maximize the throughput. In \cite{L. Sboui}, an multiple-input multiple-output CR system was studied to maximize the achievable rate. In \cite{X. Liu}, a UAV-based CR system was proposed to improve the detection performance of spectrum sensing and access the idle spectrum. In \cite{M. Hua}, a UAV-aided CR satellite terrestrial network was investigated. By jointly optimizing the base station/UAV transmit power and the UAV's trajectory, the achievable rate of the ground user can be maximized. In \cite{A. M. Almasoud}, the authors deployed a cognitive UAV to convey data to a set of Internet-of-Things devices. Although the above literature discussed the application of cognitive UAVs in various scenarios, they did not address the robust UAV's trajectory and transmit power design issues in UAV-enabled CR networks when the locations of ground nodes cannot be accurately estimated.

\begin{spacing}{-1}
\end{spacing}

\subsection{Contributions and Organization}

In this paper, we focus on designing robust trajectory and transmit power for secure UAV-enabled CR networks under two practical cases that the locations of Eves and PUs cannot be accurately obtained. Two practical estimation error models are considered, namely, the bounded location error model and the probabilistic location error model. There is one SU, multiple Eves, and multiple PUs in the networks. It is assumed that the UAV knows the accurate location of the SU, but only knows the approximate regions where PUs and Eves are located. Our goal is to maximize the average secrecy rate subject to the UAV's maximum mobility constraint, the UAV's transmit power constraint, and the average interference power constraint. To the best of our knowledge, this is the first work that studies the robust design of secure UAV-enabled CR networks under these two location error models. Although the bounded location error model has been studied in \cite{M. Cui}, \cite{X. Sun}, and the probabilistic location error model has been studied in \cite{X. Zhou}, the proposed schemes in \cite{M. Cui}, \cite{X. Sun} are only applicable to conventional wireless terrestrial communication networks and cannot work in UAV-enabled CR networks, and the physical-layer security issue in CR networks has not been considered in \cite{X. Zhou}, and thus the security of the UAV-enabled network cannot be guaranteed. Different from existing works, the main contributions of this paper are summarized as follows.

\begin{itemize}
  \item The robust trajectory and transmit power design problem is first studied in the secure UAV-enabled CR networks under the bounded location error model. The considered worst-case robust average secrecy rate maximization (WCR-ASRM) problem is intractable due to its non-convexity and the existence of semi-infinite constraints. To tackle the intractability, we propose a suboptimal algorithm to iteratively solve an approximation problem obtained by using the successive convex approximation (SCA) method and $\mathcal{S}$-Procedure.
  \item An outage-constrained robust average secrecy rate maximization (OCR-ASRM) problem is formulated for secure UAV-enabled CR networks under the probabilistic location error model. The relationship between the two location error models is studied and highlighted. The Bernstein-type inequalities are exploited to approximate the probabilistic constraints which are difficult to handle since they have no closed-from expressions. The original problem is tackled by an iterative algorithm based on the SCA method and a suboptimal solution is obtained.
  \item Simulation results and theoretical derivations show that the robust trajectory and transmit power design under the probabilistic location error model achieves a higher average secrecy rate and a lower algorithm complexity compared to that under the bounded location error model. Besides, our proposed robust schemes can achieve a larger average secrecy rate compared to benchmark schemes.
\end{itemize}

The rest of this paper is organized as follows. In Section II, the system model is presented. Section III presents a robust trajectory and transmit power design problem under the bounded location error model. Section IV presents a robust trajectory and transmit power design problem under the probabilistic location error model. In Section V, simulation results are presented. Finally, this paper is concluded in Section VI.

\emph{Notations}: Boldface capital letters and boldface lower case letters represent matrixes and vectors, respectively. ${\mathbb{C}^{M{\times}N}}$, $\mathbb{H}^{N}$, $\mathbb{R}$, and $\mathbf{I}_n$ denote the set of $M$-by-$N$ complex matrixs, the set of $N$-by-$N$ Hermitian matrices, the set of all real numbers, and a $n\times n$ identity matrix, respectively. $\mathbf{x}^{T}$ represents the transpose of a vector $\mathbf{x}$. $\Re\{\mathbf{x}\}$ denotes the real part of vector $\mathbf{x}$. $\left\| \cdot \right\|$ denotes the Euclidean norm of a vector.  $\mathbf{x}\sim\mathbb{N}(\bm{\mu},\bm{\Sigma})$ means that $\mathbf{x}$ is a real-valued random vector following a Gaussian distribution with mean $\bm{\mu}$ and covariance matrix $\bm{\Sigma}$.  $[a]^+$ denotes the function $\max(a,0)$. ${{\nabla }^{2}}f\left( x,y \right)$ denotes a square matrix composed of second-order partial derivatives of the multivariate function $f\left( x,y \right)$. $\mathbf{A}{\succeq}{\mathbf{0}}$ means that $\mathbf{A}$ is a Hermitian positive semi-definite matrix. $\text{Tr}(\mathbf{A})$ and $\text{vec}(\mathbf{A})$ denote the trace operation and the vectorization, respectively.

\section{System Model}
A downlink secure UAV-enabled CR network under spectrum sharing is considered in Fig. 1, where a cognitive UAV transmits a piece of confidential information to a ground SU over the spectrum band of the PUs, in the presence of a set of $K$ ground Eves, and $L$ ground PUs. Let $\mathcal{K}\triangleq\left\{ 1,...,K \right\}$ denote the set of Eves and $\mathcal{L}\triangleq\left\{ 1,...,L \right\}$ denote the set of PUs. In this paper, we consider a three-dimensional Cartesian coordinate system where the SU, the $l$th PU, and the $k$th Eve, $l\in \mathcal{L}$, $k\in \mathcal{K}$, have fixed  horizontal location of ${{\mathbf{q}}_{s}}=\left( {{x}_{s}},{{y}_{s}} \right)^T$, ${{\mathbf{q}}_{l}}=\left( {{x}_{p,l}},{{y}_{p,l}} \right)^T$, and ${{\mathbf{e}}_{k}}=\left( {{x}_{e,k}},{{y}_{e,k}} \right)^T$, respectively. Similar to the work in \cite{F. Zhou2}, it is assumed that the UAV flies at a constant altitude $H$ to avoid encountering obstacles. Specifically, in the horizontal direction, the UAV flies from the pre-determined initial location ${{\mathbf{q}}_{I}}=\left( {{x}_{I}},{{y}_{I}} \right)^T$ to the final location ${{\mathbf{q}}_{F}}=\left( {{x}_{F}},{{y}_{F}} \right)^T$ with the time-varying horizontal location $\mathbf{q}\left( t \right)=\left( x\left( t \right),y\left( t \right) \right)^T$, where $t\in \mathcal{T}=[0,T]$ and $T$ is the flight duration.

\begin{figure}[!t]
\centering
\includegraphics[width=3.5 in]{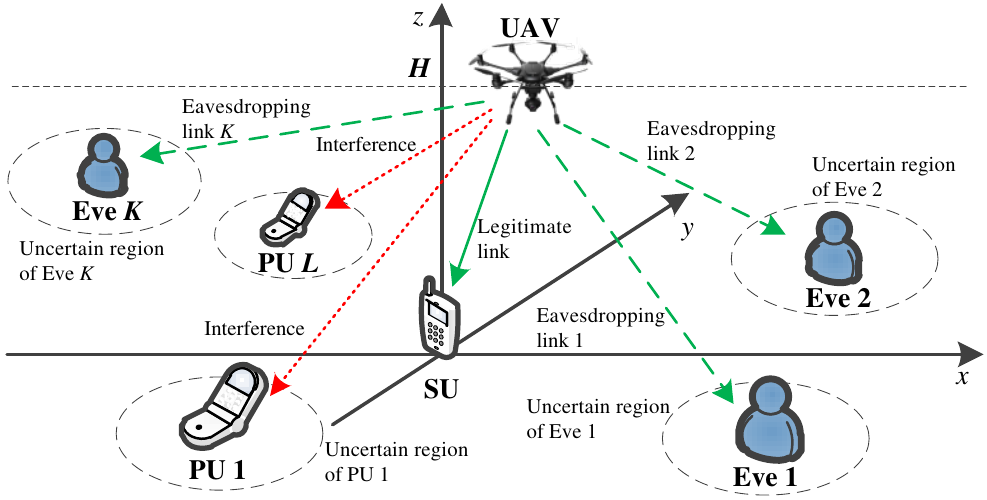}
\caption{A UAV-enabled CR system with one SU, $K$ Eves, and $L$ PUs.}
\label{fig.1}
\end{figure}

In order to facilitate the trajectory design of the UAV, the flight duration $T$ is equally divided into $N$ sufficiently small time slot with each duration of $d_t$. Due to the sufficiently small $d_t$, the state of the UAV in each time slot can be regarded as static. Thus, the flight trajectory of the UAV can be expressed as $\mathbf{q}\left[ n \right]=\left( x\left[ n \right],y\left[ n \right] \right)^T$, where $n\in \mathcal{N}\triangleq \left\{ 1,...,N \right\}$. The initial location ${{\mathbf{q}}_{I}}$ and the final location ${{\mathbf{q}}_{F}}$ can be denoted as $\mathbf{q}\left[ 1 \right]$ and $\mathbf{q}\left[ N \right]$, respectively. Considering a practical speed constraint of the UAV, the mobility constraint of the UAV can be expressed as
\begin{align}
&{{\left\| \mathbf{q}\left[ n \right]-\mathbf{q}\left[ n-1 \right] \right\|}^{2}}\le {{\left( {{V}_{\max }}{{d}_{t}} \right)}^{2}}\text{,  }\ \forall n\in \mathcal{N},
\end{align}
where ${{V}_{\max }}$ is the maximum available horizontal speed of the UAV. Let $P[n]$ denote the transmit power of the UAV in time slot $n$, $\bar{P}$ and ${P}_{\max}$ denote the average transmit power and the peak transmit power of the UAV, respectively. Thus, the average and peak transmit power constraints imposed on the UAV can be given as, respectively,
{\begin{subequations}
\begin{align}
&\frac{1}{N}\sum\limits_{n=1}^{N}{P\left[ n \right]}\le \bar{P}, \\
&\text{  0}\le P\left[ n \right]\le {{P}_{\max }}\text{,  }\ \forall n\in \mathcal{N}.
\end{align}
\end{subequations}

Since UAVs generally operate above moderate altitude \cite{X. Lin}, the channels from the UAV to SU, PUs, and Eves are all dominated by LoS components \cite{M. Cui}, \cite{X. Sun}. Thus, the distance between the UAV and the receivers can be used to determine the corresponding channels. Specifically, the channel power gain from the UAV to the SU in time slot $n$ is given by
\begin{align}
&h\left[ n \right]=\frac{{{\beta }_{0}}}{{{\left\| \mathbf{q}\left[ n \right]-{{\mathbf{q}}_{s}} \right\|}^{2}}+{{H}^{2}}},
\end{align}
where ${\beta }_{0}$ denotes the channel power gain with reference distance 1 $\text{m}$. In time slot $n$, the channel power gains from the UAV to the $l$th PU and the $k$th Eve are given as, respectively,
\begin{align}
&s_{l}\left[ n \right]=\frac{{{\beta }_{0}}}{{{\left\| \mathbf{q}\left[ n \right]-\mathbf{q}_l \right\|}^{2}}+{{H}^{2}}},
\end{align}
\begin{spacing}{0.4}
\end{spacing}
\begin{align}
&{{g}_{k}}\left[ n \right]=\frac{{{\beta }_{0}}}{{{\left\| \mathbf{q}\left[ n \right]-{{\mathbf{e}}_{k}} \right\|}^{2}}+{{H}^{2}}}.
\end{align}
Note that accurate $\mathbf{q}_l$ and $\mathbf{e}_k$ are unknown due to the existence of location estimation errors and quantization errors. It is assumed that the potential location regions of the PUs and Eves can be estimated at the UAV. This assumption can be justified from two aspects. On the one hand, due to the LoS link in UAV communications, it is possible for a UAV to obtain the approximate location of eavesdroppers via a camera or a synthetic aperture radar [35]. On the other hand, the UAV can obtain the estimated location of the PUs via cooperation between the primary network and the secondary network [27]. Moreover, Eves may be legitimate SUs in the past but do not have the privilege to access the confidential information in the current communication slots [21], [22]. In this case, the location of Eves can be estimated via the previous cooperation between Eves and the UAV.

In order to guarantee the QoS of all PUs, the interference temperature (IT) constraint is applied \cite{R. Zhang}. In this case, the interference caused by the UAV to all PUs is controlled and limited. The average interference power constraint of the $l$th PU is given by
\begin{align}
&\frac{1}{N}\sum\limits_{n=1}^{N}{s_{l}\left[ n \right]P\left[ n \right]}\le {\Upsilon},\ \ l\in \mathcal{L},
\end{align}
where ${\Upsilon}$ denotes the constant IT threshold imposed on all PUs. On the other hand, to guarantee secure transmission from the UAV to the SU, the physical layer security technique is employed, and the average secrecy rate of the SU during the flight time $T$ is given by
{\begin{subequations}
\begin{align}
&R^{sec}=\frac{1}{N}\sum\limits_{n=1}^{N}{\underset{k\in \mathcal{K}}{\mathop{\min }}\,\left\{ {{\big[ {{R}_{U}}\left[ n \right]-{{R}_{E,k}}\left[ n \right] \big]}^{+}} \right\}},\\
&{{R}_{U}}\left[ n \right]={{\log }_{2}}\left( 1+\frac{P\left[ n \right]h\left[ n \right]}{{{\sigma }^{2}}} \right),\\
&{{R}_{E,k}}\left[ n \right]={{\log }_{2}}\left( 1+\frac{P\left[ n \right]{{g}_{k}}\left[ n \right]}{\sigma _{k}^{2}} \right),
\end{align}
\end{subequations}
where ${\sigma }^{2}$ and ${\sigma _{k}^{2}}$ denote the variance of Gaussian noises at the SU receiver and the $k$th Eve receiver, respectively.

In this paper, since the perfect location of PUs and Eves are unknown, two robust trajectory and transmit power designs are proposed, namely, the worst case and the outage-constrained case. The details are presented as follows.

\section{Worst-Case Robust Trajectory and Transmit Power Design}
In this section, the robust trajectory and transmit power design problem is considered under the bounded location error model. Under this practical location uncertainty model, a lower bounded performance of UAV-enabled CR networks can be estimated and guaranteed. Moreover, this model is appropriate in many military applications requiring a high-level of security. In this work, the average secrecy rate is maximized subject to the mobility constraint, the transmit power constraint, and the average interference power constraint. Since the robust trajectory and transmit power design problem is non-convex with infinite inequality constraints, a suboptimal iterative algorithm is proposed based on the SCA method and $\mathcal{S}$-Procedure. In each iteration, a convex optimization problem is solved by using the standard convex optimization method.

\begin{spacing}{-0.45}
\end{spacing}

\subsection{The Bounded Location Error Model}
According to the works in \cite{M. Cui} and \cite{X. Sun}, in practice, the estimation errors for the location of PUs and Eves can be modeled as the bounded location error model. The bounded location error model for the location of the $l$th PU is given as
{\begin{subequations}
\begin{align}
{{\mathbf{q}}_{l}}&={{\mathbf{\bar{q}}}_{l}}+\Delta {{\mathbf{q}}_{l}},\ \ l\in \mathcal{L},\\
{{\Psi }_{l}}&\triangleq \left\{ \Delta {{\mathbf{q}}_{l}}\in {{\mathbb{R}}^{2\times 1}}\text{ }:\text{ }{{\left\| \Delta {{\mathbf{q}}_{l}} \right\|}^{2}}\le {{\omega}^{2}_{l}} \right\},
\end{align}
\end{subequations}
and the bounded location error model for the location of the $k$th Eve is given as
{\begin{subequations}
\begin{align}
{{\mathbf{e}}_{k}}&={{\mathbf{\bar{e}}}_{k}}+\Delta {{\mathbf{e}}_{k}},\ \ k\in \mathcal{K},\\
{{\Omega }_{k}}&\triangleq \left\{ \Delta {{\mathbf{e}}_{k}}\in {{\mathbb{R}}^{2\times 1}}\text{ }:\text{ }{{\left\| \Delta {{\mathbf{e}}_{k}} \right\|}^{2}}\le {{\xi}^{2}_{k}} \right\},
\end{align}
\end{subequations}
where ${{\mathbf{\bar{q}}}_{l}}$ and ${{\mathbf{\bar{e}}}_{k}}$ denote the estimates of the location vectors ${{\mathbf{q}}_{l}}$ and ${{\mathbf{e}}_{k}}$, respectively; $\Delta {{\mathbf{q}}_{l}}$ and $\Delta {{\mathbf{e}}_{k}}$ denote the location estimation errors of ${{\mathbf{q}}_{l}}$ and ${{\mathbf{e}}_{k}}$, respectively; ${{\Psi }_{l}}$ and ${{\Omega }_{k}}$ represent the uncertainty regions of ${{\mathbf{q}}_{l}}$ and ${{\mathbf{e}}_{k}}$, respectively; ${{\omega}_{l}}$ and ${{\xi}_{k}}$ denote the radii of the uncertainty regions ${{\Psi }_{l}}$ and ${{\Omega }_{k}}$, respectively.

\begin{spacing}{-0.45}
\end{spacing}

\subsection{Worst-Case Robust ASRM Problem}
Based on the bounded location error model, the WCR-ASRM problem subject to the UAV mobility constraint, the transmit power constraint, and the average interference power constraint is formulated as
{\begin{subequations}
\begin{align}
&\underset{\mathcal{P},\mathcal{Q}}{\mathop{\max }}\,\frac{1}{N}\sum\limits_{n=1}^{N}{\underset{k\in \mathcal{K}}{\mathop{\min }}\,\left\{ {{R}_{U}}\left[ n \right]-\underset{\Delta {{\mathbf{e}}_{k}}\in {{\Omega }_{k}}}{\mathop{\max }}\,{{R}_{E,k}}\left[ n \right] \right\}}\\
&\text{s.t.} {\quad}\ \frac{1}{N}\sum\limits_{n=1}^{N}{{{s}_{l}}\left[ n \right]P\left[ n \right]}\le {\Upsilon} ,\text{ } \forall \Delta {{\mathbf{q}}_{l}}\in {{\Psi }_{l}},\ l\in \mathcal{L},\\
&{\quad}\ {\quad}\ \text{(1)\,,\ (2a)\,,\ (2b),}
\end{align}
\end{subequations}
where $\mathcal{P}$ and $\mathcal{Q}$ denote the sets of variables $\left\{ P\left[ n \right] \right\}_{n=1}^{N}$ and $\left\{ \mathbf{q}\left[ n \right] \right\}_{n=1}^{N}$, respectively. (10b) can guarantee that the average interference power caused by the UAV to each PU does not exceed the IT threshold for all location estimation errors satisfying (8b). Owing to the nonlinear objective function and the non-convex constraint in (10b), the original optimization problem (10) is non-convex and challenging to solve. Moreover, the infinite inequality constraint caused by the uncertain region complicates the problem. Note that the operator $[\cdot]^+$ in the objective function can be safely omitted without affecting the optimal value of the optimization problem. For more relevant proof details, please refer to the work in \cite{M. Cui}.

In general, the formulated problem is non-convex and intractable. As a compromise approach, we aim to design a suboptimal algorithm to achieve an efficient solution. To this end, slack variables ${{\alpha }_{n}}$, ${{\beta }_{n}}$, and ${{\gamma }_{l,n}}$, where $n\in \mathcal{N}$ and $l\in \mathcal{L}$, are introduced. Thus, (10) can be equivalently expressed as
{\begin{subequations}
\begin{align}
&\underset{\mathcal{P},\mathcal{Q},\bm{\alpha},\bm{\beta},\bm{\gamma}}{\mathop{\max }}\,\frac{1}{N}\sum\limits_{n=1}^{N}\{{{\log }_{2}}\left( {{\alpha }_{n}} \right)-{{\beta }_{n}}\}\\
&\text{s.t.}\ \ \frac{1}{N}\sum\limits_{n=1}^{N}{{{\gamma }_{l,n}}}\le {\Upsilon},\ \forall l,\\
&\ \;{{\left\| \mathbf{q}\left[ n \right]\!-\!{{\mathbf{q}}_{s}} \right\|}^{2}}\!+\!{{H}^{2}}\!-\!\frac{{{\beta }_{0}}P\left[ n \right]}{{{\sigma }^{2}}\left( {{\alpha }_{n}}-1 \right)}\le 0,\text{ }\forall n,\\
&{{\left\| \mathbf{q}\left[ n \right]\!-\!{{\mathbf{e}}_{k}} \right\|}^{2}}\!+\!{{H}^{2}}\!-\!\frac{{{\beta }_{0}}P\left[ n \right]}{\sigma _{k}^{2}\left( {{2}^{{{\beta }_{n}}}} \!-\!1 \right)}\!\ge \!0, \forall k,\!n,\!\Delta {{\mathbf{e}}_{k}}\!\in\!{{\Omega }_{k}},\\
&{{\left\| \mathbf{q}\left[ n \right]-{{\mathbf{q}}_{l}} \right\|}^{2}}\!+\!{{H}^{2}}\!-\!\frac{{{\beta }_{0}}}{{{\gamma }_{l,n}}}P\left[ n \right]\ge 0,\ \forall l,n,\Delta {{\mathbf{q}}_{l}}\!\in\!{{\Psi }_{l}},\\
&\ \ {\alpha }_{n}>1,\text{ }\forall n,\\
&\ \ \text{(1)\,,\ (2a)\,,\ (2b),}
\end{align}
\end{subequations}
where $\bm{\alpha}$, $\bm{\beta}$, and $\bm{\gamma}$ denote the sets of slack variables $\{{\alpha }_{n}\}_{n=1}^{N}$, $\{{\beta }_{n}\}_{n=1}^{N}$, and $\{\{{\gamma}_{l,n}\}_{l=1}^{L}\}_{n=1}^{N}$, respectively. Note that (11d) and (11e) are infinite inequality constraints due to the uncertainty of the location information. In order to tackle them, the $\mathcal{S}$-Procedure is introduced as follows.

\begin{figure*}[hb]
\begin{spacing}{0}
\end{spacing}
\hrulefill \\
\begin{spacing}{-1.1}
\end{spacing}
$$\;\ {\quad}{\qquad}{\qquad}-{{\log }_{2}}\left( {{\varphi }_{n}}+1 \right)\ge -{{\log }_{2}}\left( {{{\tilde{\varphi }}}_{n}}+1 \right)-\frac{1}{\left( {{{\tilde{\varphi }}}_{n}}+1 \right)\ln 2}\left( {{\varphi }_{n}}-{{{\tilde{\varphi }}}_{n}} \right)=\Phi \left( {{\varphi }_{n}} \right){\qquad}{\qquad}{\qquad}{\qquad}{\qquad}{\qquad}{\quad}\ \text{(16a)}$$\\
\begin{spacing}{-2.1}
\end{spacing}
$${\qquad}{\qquad}{\qquad}\frac{{{\beta }_{0}}}{{{\sigma }^{2}}\tau \left[ n \right]\left( {{\alpha }_{n}}-1 \right)}\ge \left( \frac{{{\beta }_{0}}}{{{\sigma }^{2}}} \right)\left( \frac{1}{\tilde{\tau }\left[ n \right]\left( {{{\tilde{\alpha }}}_{n}}-1 \right)}-\frac{\left( \tau \left[ n \right]-\tilde{\tau }\left[ n \right] \right)}{\tilde{\tau }{{\left[ n \right]}^{2}}\left( {{{\tilde{\alpha }}}_{n}}-1 \right)}-\frac{\left( {{\alpha }_{n}}-{{{\tilde{\alpha }}}_{n}} \right)}{\tilde{\tau }\left[ n \right]{{\left( {{{\tilde{\alpha }}}_{n}}-1 \right)}^{2}}} \right)=\Theta \left( \tau \left[ n \right],{{\alpha }_{n}} \right){\qquad}\ \,\text{(16b)}$$\\
\begin{spacing}{-2.0}
\end{spacing}
$${\,}{\qquad}{\qquad}{\qquad}{{\left( \mathbf{q}\left[ n \right]-{{{\mathbf{\bar{e}}}}_{k}} \right)}^{T}}\left( \mathbf{q}\left[ n \right]-{{{\mathbf{\bar{e}}}}_{k}} \right)\ge {{\left( \mathbf{\tilde{q}}\left[ n \right]-{{{\mathbf{\bar{e}}}}_{k}} \right)}^{T}}\left( 2\mathbf{q}\left[ n \right]-{{{\mathbf{\bar{e}}}}_{k}}-\mathbf{\tilde{q}}\left[ n \right] \right)
{\qquad}{\qquad}{\qquad}{\qquad}{\qquad}{\qquad}{\qquad}{\qquad}\ \ \text{(16c)}$$\\
\begin{spacing}{-2.6}
\end{spacing}
$${\,}{\qquad}{\qquad}{\qquad}{{\left( \mathbf{q}\left[ n \right]-{{{\mathbf{\bar{q}}}}_{l}} \right)}^{T}}\left( \mathbf{q}\left[ n \right]-{{{\mathbf{\bar{q}}}}_{l}} \right)\ge {{\left( \mathbf{\tilde{q}}\left[ n \right]-{{{\mathbf{\bar{q}}}}_{l}} \right)}^{T}}\left( 2\mathbf{q}\left[ n \right]-{{{\mathbf{\bar{q}}}}_{l}}-\mathbf{\tilde{q}}\left[ n \right] \right)
{\qquad}{\qquad}{\qquad}{\qquad}{\qquad}{\qquad}{\qquad}{\qquad}\ \ \ \text{(16d)}$$
\end{figure*}

\emph{\textbf{Lemma 1 ($\mathcal{S}$-Procedure) \cite{S. Boyd}:}} Let ${{f}_{j}}\left( \mathbf{x} \right)={{\mathbf{x}}^{T}}{{\mathbf{A}}_{j}}\mathbf{x}+2\Re \left\{ \mathbf{b}_{j}^{T}\mathbf{x} \right\}+{{c}_{j}}$, $j\in \left\{ 1,2 \right\}$, where $\mathbf{x}\in {\mathbb{C}^{N\times 1}}$, ${{\mathbf{A}}_{j}}\in {{\mathbb{H}}^{N}}$, ${{\mathbf{b}}_{j}}\in {\mathbb{C}^{N\times 1}}$, and ${{c}_{j}}\in \mathbb{R}$. Then, the expression ${{f}_{1}}\left( \mathbf{x} \right)\le 0\Rightarrow {{f}_{2}}\left( \mathbf{x} \right)\le 0$ holds if and only if there exists a $\mu \ge \text{0}$ such that
\begin{align}
&\mu \left[ \begin{matrix}
   {{\mathbf{A}}_{1}} & {{\mathbf{b}}_{1}}  \\
   \mathbf{b}_{1}^{H} & {{c}_{1}}  \\
\end{matrix} \right]-\left[ \begin{matrix}
   {{\mathbf{A}}_{2}} & {{\mathbf{b}}_{2}}  \\
   \mathbf{b}_{2}^{H} & {{c}_{2}}  \\
\end{matrix} \right]\succeq \mathbf{0},
\end{align}
provided that there exists a vector ${\mathbf{\hat{x}}}$ such that ${{f}_{\text{1}}}\left( {\mathbf{\hat{x}}} \right)<0$.

By applying the $\mathcal{S}$-Procedure and introducing slack variables $\theta_{k,n}$, $\chi_{l,n}$, $\lambda_{k,n}$, and $\mu_{l,n}$, where $n\in \mathcal{N}$, $l\in \mathcal{L}$, and $k\in \mathcal{K}$, (11) can be expressed equivalently as
{\begin{subequations}
\begin{align}
&\underset{\mathcal{P},\mathcal{Q},\bm{\alpha},\bm{\beta},\bm{\gamma},\bm{\theta},\bm{\chi},\bm{\lambda},\bm{\mu}}{\mathop{\max }}\,\frac{1}{N}\sum\limits_{n=1}^{N}\{{{\log }_{2}}\left( {{\alpha }_{n}} \right)-{{\beta }_{n}}\}\\
&\text{s.t.}\ \ \left[ \begin{matrix}
   \left( {{\lambda }_{k,n}}+1 \right){{\mathbf{I}}_{2}} & -\left( \mathbf{q}\left[ n \right]-{{{\mathbf{\bar{e}}}}_{k}} \right)  \\
   -{{\left( \mathbf{q}\left[ n \right]-{{{\mathbf{\bar{e}}}}_{k}} \right)}^{T}} & {{c}_{k,n}}-{{\lambda }_{k,n}}\xi _{k}^{2}  \\
\end{matrix} \right]\succeq \mathbf{0},\ \forall k,n,\\
&{\qquad}\frac{{{\beta }_{0}}P\left[ n \right]}{\sigma _{k}^{2}\left( {{2}^{{{\beta }_{n}}}}-1 \right)}\le {{\theta }_{k,n}},\ \forall k,n,\\
&{\qquad}\left[ \begin{matrix}
   \left( {{\mu }_{l,n}}+1 \right){{\mathbf{I}}_{2}} & -\left( \mathbf{q}\left[ n \right]-{{{\mathbf{\bar{q}}}}_{l}} \right)  \\
   -{{\left( \mathbf{q}\left[ n \right]-{{{\mathbf{\bar{q}}}}_{l}} \right)}^{T}} & {{c}_{l,n}}-{{\mu }_{l,n}}\omega_{l}^{2}  \\
\end{matrix} \right]\succeq \mathbf{0},\text{ }\forall l,n,\\
&{\qquad}\frac{{{\beta }_{0}}}{{{\gamma }_{l,n}}}P\left[ n \right]\le {{\chi }_{l,n}},\text{ }\forall l,n,\\
&{\qquad}{{\lambda}_{k,n}}\geq 0,\ \forall k,n,\ \ {\mu }_{l,n}\geq 0,\ \forall l,n,\\
&{\qquad}\text{(11b)\,,\ (11c)\,,\ (11f)\,,\ (11g),}
\end{align}
\end{subequations}
where ${{c}_{k,n}}$ and ${{c}_{l,n}}$ are given as, respectively,
{\begin{subequations}
\begin{align}
&{{c}_{k,n}}={{\left( \mathbf{q}\left[ n \right]-{{{\mathbf{\bar{e}}}}_{k}} \right)}^{T}}\left( \mathbf{q}\left[ n \right]-{{{\mathbf{\bar{e}}}}_{k}} \right)+{{H}^{2}}-{{\theta }_{k,n}},\\
&{{c}_{l,n}}={{\left( \mathbf{q}\left[ n \right]-{{{\mathbf{\bar{q}}}}_{l}} \right)}^{T}}\left( \mathbf{q}\left[ n \right]-{{{\mathbf{\bar{q}}}}_{l}} \right)+{{H}^{2}}-{{\chi }_{l,n}},
\end{align}
\end{subequations}
where $\bm{\theta}$, $\bm{\chi}$, $\bm{\lambda}$, and $\bm{\mu}$ denote the sets of slack variables $\{\{\theta_{k,n}\}_{k=1}^{K}\}_{n=1}^{N}$, $\{\{\chi_{l,n}\}_{l=1}^{L}\}_{n=1}^{N}$, $\{\{\lambda_{k,n}\}_{k=1}^{K}\}_{n=1}^{N}$, and $\{\{{\mu}_{l,n}\}_{l=1}^{L}\}_{n=1}^{N}$, respectively. Thus, there are finite numbers of inequality constraints in problem (13). However, (13) is still non-convex due to the existence of variables coupling in (11c), (13b), (13c), (13d), and (13e). To solve this difficulty, we introduce the slack variables, $\tau \left[ n \right]={{\left( P\left[ n \right] \right)}^{-1}}$ and ${{\varphi }_{n}}= {{2}^{{{\beta }_{n}}}}-1$, where $n\in \mathcal{N}$. Then, the problem in (13) can be rewritten as
{\begin{subequations}
\begin{align}
&\underset{\bm{\tau},\mathcal{Q},\bm{\alpha},\bm{\varphi},\bm{\gamma},\bm{\theta},\bm{\chi},\bm{\lambda},\bm{\mu}}{\mathop{\max }}\,\frac{1}{N}\sum\limits_{n=1}^{N}\{{{\log }_{2}}\left( {{\alpha }_{n}} \right)-{{\log }_{2}}\left( {{\varphi }_{n}}+1 \right)\}\\
&\text{s.t.}\ \ \frac{{{\beta }_{0}}}{\sigma _{k}^{2}\tau \left[ n \right]{{\varphi }_{n}}}\le {{\theta }_{k,n}},\text{ }\forall k,n,\\
&{\qquad}\frac{{{\beta }_{0}}}{\tau \left[ n \right]{{\gamma }_{l,n}}}\le {{\chi }_{n}},\text{ }\forall l,n,\\
&{\quad}{\quad}{{\left\| \mathbf{q}\left[ n \right]\!-\!{{\mathbf{q}}_{s}} \right\|}^{2}}\!+\!{{H}^{2}}\!-\!\frac{{{\beta }_{0}}}{{{\sigma }^{2}}\tau \left[ n \right]\left( {{\alpha }_{n}}\!-\!1 \right)}\le 0,\text{  }\forall n,\\
&{\qquad}\frac{1}{N}\sum\limits_{n=1}^{N}{\frac{1}{\tau \left[ n \right]}}\le \bar{P},\ \tau \left[ n \right]\ge \frac{1}{{{P}_{\max }}},\text{  }\forall n,\\
&{\qquad}\text{(1)\,,\ (11b)\,,\ (11f)\,,\ (13b)\,,\ (13d)\,,\ (13f)},
\end{align}
\end{subequations}
where $\bm{\tau}$ and $\bm{\varphi}$ denote the sets of the slack variables $\{\tau_{n}\}_{n=1}^{N}$ and $\{\varphi_{n}\}_{n=1}^{N}$, respectively. Note that (15b) and (15c) are convex constraints. However, (15d) is a non-convex constraint which is verified in the following proposition.

\emph{\textbf{Proposition 1 :}} Let $f\left( x,y \right)\triangleq \frac{1}{xy}$, where $x>0$ and $y>0$. Then, $f\left( x,y \right)$ is jointly convex with respect to $x$ and $y$.

\begin{IEEEproof}
Please refer to Appendix A.
\end{IEEEproof}

\begin{figure*}[hb]
\begin{spacing}{0}
\end{spacing}
\hrulefill \\
\begin{spacing}{-1.1}
\end{spacing}
$$\,\mathcal{O}\left({{T}_{1}}\sqrt{\left( 5K+5L+4 \right)N+L+1}\ln \left( {{\epsilon }^{-1}} \right)m \Big\{ \left( {27}+{9}m \right)\left( K+L \right)N+\big[ 2\left( K+L+2 \right)N+L+1 \big]\left( 1+m \right)+{{m}^{2}} \Big\}\right)
 \, \text{(19)}$$
\end{figure*}

Meanwhile, the objective function in (15a) is non-concave, since $-{{\log }_{2}}\left( {{\varphi }_{n}}+1 \right)$ is convex. Moreover, constraints (13b) and (13d) are non-convex due to the existence of the second-order variables $x^2[n]$ and $y^2[n]$ in $c_{k,n}$ and $c_{l,n}$. Fortunately, the above functions are convex which are lower bounded by the corresponding first-order Taylor approximations. For a given set of feasible points, $\left\{ \tilde{\tau }\left[ n \right],\mathbf{\tilde{q}}\left[ n \right],{{{\tilde{\alpha }}}_{n}},{{{\tilde{\varphi }}}_{n}} \right\}_{n=1}^{N}$, inequalities (16) shown at the bottom of this page must hold $\forall n\in \mathcal{N}$. Then, the lower bound value of (15) can be obtained by solving the following problem:
\setcounter{equation}{16}
{\begin{subequations}
\begin{align}
&\underset{\bm{\tau},\mathcal{Q},\bm{\alpha},\bm{\varphi},\bm{\gamma},\bm{\theta},\bm{\chi},\bm{\lambda},\bm{\mu}}{\mathop{\max }}\,\frac{1}{N}\sum\limits_{n=1}^{N}\{{{\log }_{2}}\left( {{\alpha }_{n}} \right)+\Phi \left( {{\varphi }_{n}} \right)\}\\
&\text{s.t.}\ \ \ {{\left\| \mathbf{q}\left[ n \right]-{{\mathbf{q}}_{s}} \right\|}^{2}}+{{H}^{2}}-\Theta \left( \tau \left[ n \right],{{\alpha }_{n}} \right)\le 0,\text{  }\forall n,\\
&{\qquad}\left[ \begin{matrix}
   \left( {{\lambda }_{k,n}}+1 \right){{\mathbf{I}}_{2}} & -\left( \mathbf{q}\left[ n \right]-{{{\mathbf{\bar{e}}}}_{k}} \right)  \\
   -{{\left( \mathbf{q}\left[ n \right]-{{{\mathbf{\bar{e}}}}_{k}} \right)}^{T}} & {\tilde{c}_{k,n}}-{{\lambda }_{k,n}}\xi _{k}^{2}  \\
\end{matrix} \right]\succeq \mathbf{0},\ \forall k,n,\\
&{\qquad}\left[ \begin{matrix}
   \left( {{\mu }_{l,n}}+1 \right){{\mathbf{I}}_{2}} & -\left( \mathbf{q}\left[ n \right]-{{{\mathbf{\bar{q}}}}_{l}} \right)  \\
   -{{\left( \mathbf{q}\left[ n \right]-{{{\mathbf{\bar{q}}}}_{l}} \right)}^{T}} & {\tilde{c}_{l,n}}-{{\mu }_{l,n}}\omega _{l}^{2}  \\
\end{matrix} \right]\succeq \mathbf{0},\text{ }\forall l,n,\\
&{\qquad}\ \text{(1)\,,\ (11b)\,,\ (11f)\,,\ (13f)\,,\ (15b)\,,\ (15c)\,,\ (15e),}
\end{align}
\end{subequations}
where $\tilde{c}_{k,n}$ an $\tilde{c}_{l,n}$ are respectively given as, respectively,
{\begin{subequations}
\begin{align}
{\tilde{c}_{k,n}}&={{\left( \mathbf{\tilde{q}}\left[ n \right]-{{{\mathbf{\bar{e}}}}_{k}} \right)}^{T}}\!\left( 2\mathbf{q}\left[ n \right]\!-{{{\mathbf{\bar{e}}}}_{k}}\!-\mathbf{\tilde{q}}\left[ n \right] \right)\!+\!{{H}^{2}}\!-\!{{\theta }_{k,n}},\\
{\tilde{c}_{l,n}}&={{\left( \mathbf{\tilde{q}}\left[ n \right]-{{{\mathbf{\bar{q}}}}_{l}} \right)}^{T}}\!\left( 2\mathbf{q}\left[ n \right]\!-{{{\mathbf{\bar{q}}}}_{l}}\!-\mathbf{\tilde{q}}\left[ n \right] \right)\!+\!{{H}^{2}}\!-\!{{\chi }_{l,n}}.
\end{align}
\end{subequations}
Note that (17) is a convex optimization problem, which can be efficiently solved by using the standard convex optimization method, e.g., the interior-point method \cite{S. Boyd}. Then, Algorithm 1 based on the SCA method is proposed to tighten the first-order Taylor approximations. In this case, the suboptimal solutions of the original optimization problem (10), $\mathcal{P}^{*}$ and $\mathcal{Q}^{*}$, can be obtained by iteratively solving (17).

\begin{tabular}{lcl}
\\\toprule
\!$\textbf{Algorithm 1}$: Iterative Algorithm for solving (10) \\ \midrule
\;1: \textbf{Initialization:}\\
\;\ \ \ \ \ Initialize $\left\{ \tilde{\tau }\left[ n \right],\mathbf{\tilde{q}}\left[ n \right],{{{\tilde{\alpha }}}_{n}},{{{\tilde{\varphi }}}_{n}} \right\}_{n=1}^{N}$, \\
\;\ \ \ \ \ the threshold $\epsilon\rightarrow0$.\\
\;2: \textbf{Optimization:}\\
\;\ \ \ \ \ \textbf{repeat} \\
\;\ \ \ \ \ \ \ 1) Obtain suboptimal $\bm{\tau}^{*}$, $\mathcal{Q}^{*}$, $\bm{\alpha}^{*}$, and $\bm{\varphi}^{*}$ by\\
\;\ \ \ \ \ \ \ \ \,\,\, solving (17). \\
\;\ \ \ \ \ \ \ 2) if $\left| R^{sec}\left( {{\mathcal{P}}^{*}},{{\mathcal{Q}}^{*}} \right)-R^{sec}( \tilde{\mathcal{P}},\tilde{\mathcal{Q}} ) \right|\le\epsilon$\\
\;\ \ \ \ \ \ \ \ \ \ \,\,\, \textbf{break};\\
\;\ \ \ \ \ \ \ \ \,\, end\\
\;\ \ \ \ \ \ \ 3) Update $\{\tilde{\tau }\left[ n \right] \}_{n=1}^{N}\leftarrow\bm{\tau}^{*}$, $\{\mathbf{\tilde{q}}\left[ n \right]\}_{n=1}^{N}\leftarrow\mathcal{Q}^{*}$,  \\
\;\ \ \ \ \ \ \ \ \,\,\,\,\, {\quad}{\qquad}$\{ {{{\tilde{\alpha }}}_{n}} \}_{n=1}^{N}\leftarrow\bm{\alpha}^{*}$, $\{ {{{\tilde{\varphi }}}_{n}} \}_{n=1}^{N}\leftarrow\bm{\varphi}^{*}$.\\
\;3: \textbf{Output} $\lbrace\mathcal{P}^{*}\!$ and $\mathcal{Q}^{*}\rbrace$\\
\bottomrule
\end{tabular}
\vspace*{10pt}

\emph{\textbf{Remark 1:}} Due to the inequalities in (16b), (16c), and (16d), the left hand side (LHS) of (15d) is a lower bound of (17b), and the LHS of (13b) and (13d) are upper bounds for (17c) and (17d), respectively. In other words, (17b), (17c), and (17d) hold implying (15d), (13b), and (13d) hold, respectively. Hence, the solution of problem (17) is a feasible suboptimal solution of problem (15).

\emph{\textbf{Remark 2:}} The convergence of Algorithm 1 is guaranteed. Since (16a) is a lower bound of $-{{\log }_{2}}\left( {{\varphi }_{n}}+1 \right)$, the objective function of (17) is a lower bound of the objective function of (15). Note that the objective value of (15) is equal to that of (17) only at feasible points $\left\{ \tilde{\tau }\left[ n \right],\mathbf{\tilde{q}}\left[ n \right],{{{\tilde{\alpha }}}_{n}},{{{\tilde{\varphi }}}_{n}} \right\}_{n=1}^{N}$, and the objective value of (15) is larger than that of (17) with the solution of (17). Thus, the objective value of (15) with the solution of (17) is no less than that with the solution $\left\{ \tilde{\tau }\left[ n \right],\mathbf{\tilde{q}}\left[ n \right],{{{\tilde{\alpha }}}_{n}},{{{\tilde{\varphi }}}_{n}} \right\}_{n=1}^{N}$. It means that the objective value of (17) and (15) is non-descending over iteration.

\emph{\textbf{Remark 3:}} In step 1 of Algorithm 1, the convex optimization problem in (17) can be solved by using the interior-point method (IPM) \cite{S. Boyd}. According to \cite{A. Ben-Tal}, the computational complexity of Algorithm 1 using the IPM can be divided into three parts, namely, the required numbers of iterations of the SCA method, the iteration complexity, and the per-iteration computation cost. It is assumed that the maximum number of iterations of the SCA method is $T_1$ and the accuracy of the iteration is $\epsilon$. Problem (17) has $(K+L)N$ linear matrix inequalities (LMIs) of size 3, $2(K+L+2)N+L+1$ LMIs of size 1. The number of decision variables $m$ is on the order of $(K+L)N$, e.g., $m=\mathcal{O}((K+L)N)$, where $\mathcal{O}(\cdot)$ is the big-O notation. Thus, the total complexity of Algorithm 1 under the bounded location error model is given by (19), as shown at the bottom of this page. Note that the proposed Algorithm 1 has a polynomial time computational complexity which is suitable for practical implementation.

\section{Outage-Constrained Robust Trajectory and Transmit Power Design}
In this section, the robust trajectory and transmit power design problem is considered under the probabilistic location error model. The probabilistic location error model is different from the bounded location error model as the latter is appropriate to handle the extreme scenario where the system performance has to be guaranteed even in the worst case. Under the bounded location error model, the performance is usually conservative since the worst-case objective function is considered and all constraints should be rigorously hold. As an alternative, the probabilistic location error model is considered for tackling the issues caused by over protection. In this case, the constraints involving the location error model are outage probability constraints. Note that the OCR-ASRM problem is even more challenging compare to the WCR-ASRM problem due to the existence of probability constraints. Furthermore, to provide a fair comparison between the performance under two different location error models, a method is proposed to ensure that the WCR-ASRM problem is a safe approximation to the OCR-ASRM problem \cite{Q. Li2}. In this case, each feasible point of the WCR-ASRM problem is also feasible to the OCR-ASRM problem, and always satisfies the corresponding outage probability constraints.

\begin{spacing}{-0.6}
\end{spacing}

\subsection{The Probabilistic Location Error Model}
According to the work in \cite{X. Zhou}, it is assumed that the locations of each PU and each Eve are stochastic and follow the Gaussian distribution due to the estimation errors. Specifically, $x$ and $y$ coordinates are independent and identically distributed (i.i.d.) Gaussian random variables. The Gaussian location error models for the location of the $l$th PU and the $k$th Eve, are given as, respectively,
\setcounter{equation}{19}
{\begin{subequations}
\begin{align}
{{\mathbf{q}}_{l}}&={{\mathbf{\bar{q}}}_{l}}+\Delta {{\mathbf{q}}_{l}},\ \Delta {{\mathbf{q}}_{l}}\sim \mathbb{N}\left( \mathbf{0},{{\varpi}^{2}_{l}}\mathbf{I}_2 \right),\ \forall l\in \mathcal{L},\\
{{\mathbf{e}}_{k}}&={{\mathbf{\bar{e}}}_{k}}+\Delta {{\mathbf{e}}_{k}},\ \Delta {{\mathbf{e}}_{k}}\sim \mathbb{N}\left( \mathbf{0},\varepsilon _{k}^{2}\mathbf{I}_2 \right),\ \forall k\in \mathcal{K},
\end{align}
\end{subequations}
where ${{\mathbf{\bar{q}}}_{l}}$ and ${{\mathbf{\bar{e}}}_{k}}$ denote the estimated location; $\Delta {{\mathbf{q}}_{l}}$ and $\Delta {{\mathbf{e}}_{k}}$ denote the location estimation errors; ${{\varpi }^{2}_{l}}$ and $\varepsilon _{k}^{2}$ are variances of the corresponding location estimation errors.

\begin{spacing}{-0.6}
\end{spacing}

\subsection{Outage-Constrained Robust ASRM Problem}
Based on the probabilistic location error model, the OCR-ASRM problem is studied under the outage probabilistic constraints which is formulated as
{\begin{subequations}
\begin{align}
&\underset{\mathcal{P},\mathcal{Q},\bm{\beta},\bm{\gamma}}{\mathop{\max }}\ \ \frac{1}{N}\sum\limits_{n=1}^{N}{\left\{ {{R}_{U}}\left[ n \right]-{{\beta }_{n}} \right\}}\\
&\text{s.t.} \ \text{P}{{\text{r}}_{\left\{ \Delta {{\mathbf{e}}_{k}} \right\}_{k=1}^{K}}}\left\{ \underset{k\in \mathcal{K}}{\mathop{\max }}\,\text{ }{{R}_{E,k}}\left[ n \right]\le {{\beta }_{n}} \right\}\ge 1-{{\rho }},\,\forall n,\\
&{\quad}\ \,\text{P}{{\text{r}}_{\left\{ \Delta {{\mathbf{q}}_{l}} \right\}}}\left\{ {{s}_{l}}\left[ n \right]P\left[ n \right]\le {{\gamma }_{l,n}} \right\}\ge 1-{{\phi }}, \ \forall l,n,\\
&{\quad}\ \,\text{(11b)\,,\ (11g),}
\end{align}
\end{subequations}
where ${\rho }\in(0,1]$ denotes the maximum outage probability associated with the transmission rate of all Eves and ${\phi }\in(0,1]$ denotes the maximum outage probability associated with the interference power of all PUs. Since the location estimation errors of Eves are probabilistic, the problem of maximizing the average secrecy rate (7a) can be transformed into a problem that maximizes (21a) while satisfying the corresponding outage probability constraint (21b) by introducing an auxiliary variable ${\beta }_{n}$. (21c) guarantees that the outage probability of the interference power of the PUs being larger than ${\gamma }_{l,n}$ is less than ${\phi }$. Note that the probability term in (21b) can be regarded as the coupling of probability terms among Eves, which is intractable and complicated. In order to decouple the joint probabilistic constraint into tractable terms, by exploiting the independence between the locations of Eves, we have the following implication:
{\begin{subequations}
\begin{align}
\text{(21b)}&\Leftrightarrow \prod\limits_{k=1}^{K}{\text{P}{{\text{r}}_{\left\{ \Delta {{\mathbf{e}}_{k}} \right\}}}\left\{ {{R}_{E,k}}\left[ n \right]\le {{\beta }_{n}} \right\}}\ge 1-\rho ,\text{ }\forall n,\\
&\Leftarrow \text{P}{{\text{r}}_{\left\{ \Delta {{\mathbf{e}}_{k}} \right\}}}\left\{ {{R}_{E,k}}\left[ n \right]\le {{\beta }_{n}} \right\}\ge 1-\bar{\rho },\text{ }\forall k,n,
\end{align}
\end{subequations}
where $\bar{\rho }=1-{{\left( 1-\rho  \right)}^{{1}/{K}}}$. (22b) means that the outage probability of mutual information for each Eve being larger than ${\beta }_{n}$ should be lower than $\bar{\rho}$. Then, by introducing the slack variables $\alpha_{n}$, $n\in\mathcal{N}$, a lower bound value of problem (21) can be obtained by solving the following problem:
{\begin{subequations}
\begin{align}
&\underset{\mathcal{P},\mathcal{Q},\bm{\alpha},\bm{\beta},\bm{\gamma}}{\mathop{\max }}\ \ \frac{1}{N}\sum\limits_{n=1}^{N}{\left\{ {{\log }_{2}}\left( {{\alpha }_{n}} \right)-{{\beta }_{n}} \right\}}\\
&\text{s.t.} \ \text{P}{{\text{r}}_{\left\{ \Delta {{\mathbf{e}}_{k}} \right\}}}\left\{ {{R}_{E,k}}\left[ n \right]\le {{\beta }_{n}} \right\}\ge 1-\bar{\rho },\text{ }\forall k,n,\\
&{\quad}\ {{\left\| \mathbf{q}\left[ n \right]\!-\!{{\mathbf{q}}_{s}} \right\|}^{2}}\!+\!{{H}^{2}}\!-\!\frac{{{\beta }_{0}}P\left[ n \right]}{{{\sigma }^{2}}\left( {{\alpha }_{n}}-1 \right)}\le 0,\text{ }\forall n,\\
&{\quad}\ \text{(11b)\,,\ (11f)\,,\ (11g)\,,\ (21c).}
\end{align}
\end{subequations}

For a fair comparison between the performance of the WCR-ASRM design and the OCR-ASRM design, the radii of uncertainty regions, ${\Psi }_{l}$ and ${\Omega }_{k}$, are chosen as \cite{Q. Li2}
{\begin{subequations}
\begin{align}
&{{\omega }_{l}}=\varpi_{l} \sqrt{F_{\chi _{2}^{2}}^{-1}\left( \left( 1-\phi  \right) \right)},\ \forall l\in \mathcal{L},\\
&{{\xi }_{k}}={{\varepsilon }_{k}}\sqrt{F_{\chi _{2}^{2}}^{-1}\left( {{\left( 1-\rho  \right)}^{{1}/{K}\;}} \right)},\text{   }\forall k\in \mathcal{K},
\end{align}
\end{subequations}
where $F_{\chi _{2}^{2}}^{-1}\left( \cdot  \right)$ denotes the inverse cumulative distribution function of a Chi-square random variable with 2 degrees of freedom. Thus, the uncertainty radius ${{\omega }_{l}}$ can satisfy $\text{P}{{\text{r}}_{\left\{ \Delta {{\mathbf{q}}_{l}} \right\}}}\left\{ {{\left\| \Delta {{\mathbf{q}}_{l}} \right\|}^{2}}\le \omega_{l}^{2} \right\}=1-\phi$ and the uncertainty radius ${{\xi }_{k}}$ can satisfy $\text{P}{{\text{r}}_{\left\{ \Delta {{\mathbf{e}}_{k}} \right\}}}\left\{ {{\left\| \Delta {{\mathbf{e}}_{k}} \right\|}^{2}}\le \xi _{k}^{2} \right\}=1-\bar{\rho }$. By choosing ${{\omega }_{l}}$ and ${{\xi }_{k}}$, if constraints (11d) and (11e) are satisfied, constraints (23b) and (21c) must be satisfied. Thus, the WCR-ASRM problem (11) can be considered as  a safe approximation to the OCR-ASRM problem (23). Note that the closed-form expressions of constraints (21c) and (23b) are difficult to derive directly due to the outage probability. To convert the probability constraints into deterministic forms, the Bernstein-type inequality is applied to provide a safe approximation of problem (23), which is summarized as follows.

\emph{\textbf{Lemma 2 (The Bernstein-type Inequality) \cite{K. Y.Wang}:}} Define $\mathbf{A}\in{{\mathbb{H}}^{N}}$, $\mathbf{x}\sim \mathbb{N}\left( \mathbf{0},\mathbf{I} \right)$, ${{\mathbf{b}}}\in {\mathbb{C}^{N\times 1}}$, ${{c}}\in \mathbb{R}$, and ${\rho}\in(0,1]$, the following implication holds
\begin{align} \notag
&\text{Pr}\left\{ {{\mathbf{x}}^{T}}\mathbf{Ax}+2\Re \left\{ {{\mathbf{x}}^{T}}\mathbf{b} \right\}+c\ge 0 \right\}\ge 1-\rho\\
&\Leftarrow \left\{ \begin{matrix}
   \text{Tr}\left( \mathbf{A} \right)-\sqrt{-2\ln \left( \rho  \right)}{{{\Upsilon} }_{1}}+\ln \left( \rho  \right){{\upsilon }_{2}}+c\ge 0,  \\
   \left\| \left[ \begin{matrix}
   \text{vec}\left( \mathbf{A} \right)  \\
   \sqrt{2}\mathbf{b}  \\
\end{matrix} \right] \right\|\le {{\upsilon }_{1}},  \\
   {{\upsilon }_{2}}{{\mathbf{I}}_{N}}+\mathbf{A}\succeq \mathbf{0},\text{   }{{\upsilon }_{2}}\ge 0 , \\
\end{matrix} \right.
\end{align}
where ${{\upsilon }_{1}}$ and ${{\upsilon }_{2}}$ are slack variables.

\begin{figure*}[hb]
\begin{spacing}{0}
\end{spacing}
\hrulefill \\
\begin{spacing}{-1.1}
\end{spacing}
$${\quad}{\quad}\ \mathcal{O}\left({{T}_{2}}\sqrt{4\left( K+L+1 \right)N+L+1}\ln \left( {{\epsilon }^{-1}} \right)m \Big\{ {49}\left( K+L \right)N+\big[ 2\left( K+L+2 \right)N+L+1 \big]\left( 1+m \right)+{{m}^{2}} \Big\}\right){\ }\,{\quad}\ \text{(33)}$$
\end{figure*}

By applying Lemma 2, an approximate expression of the outage interference power constraint (21c) can be given as
{\begin{subequations}
\begin{align}
&\text{Tr}\left( {{\mathbf{A}}_{l}} \right)\!-\!\sqrt{-2\ln \left( \phi  \right)}{{\eta }_{l,n}}\!+\!\ln \left( \phi  \right){{\zeta }_{l,n}}\!+\!{\ddot{c}_{n}}\ge 0,\ \forall l,n,\\
&\left\| \left[ \begin{matrix}
   \text{vec}\left( {{\mathbf{A}}_{l}} \right)  \\
   \sqrt{2}{{\mathbf{b}}_{n}}  \\
\end{matrix} \right] \right\|\le {{\eta }_{l,n}},\ \forall l,n,\\
&{{\zeta }_{l,n}}{{\mathbf{I}}_{2}}+{{\mathbf{A}}_{l}}\succeq \mathbf{0},\text{ }{{\zeta }_{l,n}}\ge 0,\ \forall l,n,
\end{align}
\end{subequations}
where ${{\eta }_{l,n}}$ and ${{\zeta }_{l,n}}$ are slack variables; ${{\mathbf{A}}_{l}}=\varpi_{l}^{2}{{\mathbf{I}}_{2}}$, ${{\mathbf{b}}_{n}}={{\varpi}_{l}}\left( {{{\mathbf{\bar{q}}}}_{l}}-\mathbf{q}\left[ n \right] \right)$, and
\begin{align}
{\ddot{c}_{l,n}}={{\left( \mathbf{q}\left[ n \right]-{{{\mathbf{\bar{q}}}}_{l}} \right)}^{T}}\left( \mathbf{q}\left[ n \right]-{{{\mathbf{\bar{q}}}}_{l}} \right)+{{H}^{2}}-\frac{{{\beta }_{0}}}{{{\gamma }_{l,n}}}P\left[ n \right].
\end{align}
Note that the first part of (26c) always holds since ${{\zeta }_{l,n}}+\varpi_{l}^{2}\ge 0$. Similar to the outage interference power constraint, the outage transmission rate constraints of eavesdroppers (23b) can be approximated as
{\begin{subequations}
\begin{align}
&\text{Tr}\left( \!{{\mathbf{A}}_{k}} \! \right) \!-\!\sqrt{\!-2\ln \left( {\bar{\rho }} \right)}{{\upsilon }_{k,n}}\!+\!\ln \left( {\bar{\rho }} \right){{\varsigma }_{k,n}}\!+\!{\ddot{c}_{k,n}}\!\ge\! 0,\forall k,n,\\
&\left\| \left[ \begin{matrix}
   \text{vec}\left( {{\mathbf{A}}_{k}} \right)  \\
   \sqrt{2}{{\mathbf{b}}_{k,n}}  \\
\end{matrix} \right] \right\|\le {{\upsilon }_{k,n}},\ \forall  k,n,\\
&{{\varsigma}_{k,n}}{{\mathbf{I}}_{2}}+{{\mathbf{A}}_{k}}\succeq \mathbf{0},\text{ }{{\varsigma}_{k,n}}\ge 0,\ \forall k,n,
\end{align}
\end{subequations}
where ${{\upsilon }_{k,n}}$ and ${{\varsigma}_{k,n}}$ are slack variables; ${{\mathbf{A}}_{k}}=\varepsilon _{k}^{2}{{\mathbf{I}}_{2}}$, ${{\mathbf{b}}_{k,n}}={{\varepsilon }_{k}}\left( {{{\mathbf{\bar{e}}}}_{k}}-\mathbf{q}\left[ n \right] \right)$, and
\begin{align}
{\ddot{c}_{k,n}}\!=\!{{\left( \mathbf{q}\left[ n \right]\!-\!{{{\mathbf{\bar{e}}}}_{k}} \right)}^{T}}\left( \mathbf{q}\left[ n \right]\!-\!{{{\mathbf{\bar{e}}}}_{k}} \right)\!+\!{{H}^{2}}\!-\!\frac{{{\beta }_{0}}P\left[ n \right]}{\sigma _{k}^{2}\left( {{2}^{\beta_{n}}}-1 \right)}.
\end{align}
Note that the first part of (28c) always holds since ${{\varsigma}_{k,n}}+\varepsilon _{k}^{2}\ge 0$. Thus, these LMIs in the first part of (26c) and (28c) can be omitted. To tackle the coupling of variables in (23c), (27), and (29), we introduce slack variables $\tau[n]={{\left( P\left[ n \right] \right)}^{-1}}$ and $\varphi_{n}={{2}^{{{\beta }_{n}}}}-1$. Thus, the approximation problem for (23) can be rewritten as
{\begin{subequations}
\begin{align}
&\underset{\bm{\tau},\mathcal{Q},\bm{\alpha},\bm{\varphi},\bm{\gamma},\bm{\eta},\bm{\zeta},\bm{\upsilon},\bm{\varsigma}}{\mathop{\max }}\ \ \frac{1}{N}\sum\limits_{n=1}^{N}{\left\{ {{\log }_{2}}\left( {{\alpha }_{n}} \right)-{{\log }_{2}}\left( {{\varphi }_{n}}+1 \right) \right\}}\\
&\text{s.t.} \ {{\left\| \mathbf{q}\left[ n \right]-{{\mathbf{q}}_{s}} \right\|}^{2}}\!+\!{{H}^{2}}\!-\!\frac{{{\beta }_{0}}}{{{\sigma }^{2}}\tau \left[ n \right]\left( {{\alpha }_{n}}-1 \right)}\le 0,\text{ }\forall n,\\
&\ \ \ \ \text{(1)\,,\ (15e)\,,\ (11b)\,,\ (11f)\,,\ (26a)-(26c)\,,\ (28a)-(28c),}
\end{align}
\end{subequations}
where $\bm{\eta}$, $\bm{\zeta}$, $\bm{\upsilon}$, and $\bm{\varsigma}$ denote the sets of slack variables $\{\{\eta_{l,n}\}_{l=1}^{L}\}_{n=1}^{N}$, $\{\{\zeta_{l,n}\}_{l=1}^{L}\}_{n=1}^{N}$, $\{\{\upsilon_{n}\}_{k=1}^{K}\}_{n=1}^{N}$ and $\{\{\varsigma_{n}\}_{k=1}^{K}\}_{n=1}^{N}$, respectively. Note that (30) is still a non-convex optimization problem and is difficult to solve due to the non-concave objective function, the non-convex constraint in (30b), and the existence of the second-order variables $x^2[n]$ and $y^2[n]$ in (26a) and (28a).

In order to solve (30), a suboptimal iterative Algorithm 2 is proposed based on the SCA method. To this end, we exploit the first-order Taylor approximation to obtain these inequalities in (16), which are derived in Section III. Thus, the lower bound value of problem (30) can be obtained by solving the following problem,
{\begin{subequations}
\begin{align}
&\underset{\bm{\tau},\mathcal{Q},\bm{\alpha},\bm{\varphi},\bm{\gamma},\bm{\eta},\bm{\zeta},\bm{\upsilon},\bm{\varsigma}}{\mathop{\max }}\ \ \frac{1}{N}\sum\limits_{n=1}^{N}{\left\{ {{\log }_{2}}\left( {{\alpha }_{n}} \right)+\Phi(\varphi_{n}) \right\}}\\
&\text{s.t.} \ {{\left\| \mathbf{q}\left[ n \right]-{{\mathbf{q}}_{s}} \right\|}^{2}}+{{H}^{2}}-\Theta \left( \tau \left[ n \right],{{\alpha }_{n}} \right)\le 0,\text{ }\forall n,\\
&\text{Tr}\left( {{\mathbf{A}}_{l}} \right)\!-\!\sqrt{-2\ln \left( \phi  \right)}{{\eta }_{l,n}}\!+\!\ln \left( \phi  \right){{\zeta }_{l,n}}\!+\!{\hat{c}_{l,n}}\ge 0,\forall l,n,\\
&\text{Tr}\left( \!{{\mathbf{A}}_{k}} \! \right) \!-\!\sqrt{\!-2\ln \left( {\bar{\rho }} \right)}{{\upsilon }_{k,n}}\!+\!\ln \left( {\bar{\rho }} \right){{\varsigma }_{k,n}}\!+\!{\hat{c}_{k,n}}\!\ge\! 0,\forall k,n,\\
&\ \text{(1)\,,\ (15e)\,,\ (11b)\,,\ (11f)\,,\ (26b)\,,\ (26c)\,,\ (28b)\,,\ (28c),}
\end{align}
\end{subequations}
where
{\begin{subequations}
\begin{align}
&\!{\hat{c}_{l,n}}\!=\!{{\left( \mathbf{\tilde{q}}\left[ n \right]\!-\!{{{\mathbf{\bar{q}}}}_{l}} \right)}^{T}}\!\!\left( 2\mathbf{q}\left[ n \right]\!-\!{{{\mathbf{\bar{q}}}}_{l}}\!-\!\mathbf{\tilde{q}}\left[ n \right] \right)\!+\!{{H}^{2}}\!-\!\frac{{{\beta }_{0}}}{\tau[n]{{\gamma }_{l,n}}},   \\
&\!{\hat{c}_{k,n}}\!=\!{{\left( \mathbf{\tilde{q}}\left[ n \right]\!-\!{{{\mathbf{\bar{e}}}}_{k}} \right)}^{T}}\!\!\left( 2\mathbf{q}\left[ n \right]\!-\!{{{\mathbf{\bar{e}}}}_{k}}\!-\!\mathbf{\tilde{q}}\left[ n \right] \right)\!+\!{{H}^{2}}\!-\!\frac{{{\beta }_{0}}}{\sigma _{k}^{2}\tau \left[ n \right]{{\varphi }_{n}}}.
\end{align}
\end{subequations}
Note that (31) is a convex optimization problem and can be solved effectively via standard convex optimization methods. Then, Algorithm 2 based on the SCA method can tighten these first-order Taylor approximation constraints in (31). Thus, the suboptimal solution of (30) can be obtained by iteratively solving (31). Note that (30) is a safe approximation to the original OCR-ASRM problem in (21). Thus, the suboptimal solution of (30) is also the suboptimal solution of (21). The pseudo code of Algorithm 2 under the probabilistic location error model is omitted since the procedure is similar to that of Algorithm 1.

\emph{\textbf{Remark 4:}} The computational complexity of Algorithm 2 also arises from three parts, namely, the required numbers of iterations of the SCA method, the iteration complexity, and the per-iteration computation cost. It is assumed that the maximum number of iterations of the SCA method of Algorithm 2 is $T_2$ and the accuracy of the iteration of Algorithm 2 is the same as that of Algorithm 1. Problem (31) has $2(K+L+2)N+L+1$ LMIs of size 1, $(K+L)N$ second-order cone constraints with dimension 7. The numbers of decision variables $m$ is on the order of $(K+L)N$, e.g., $m=\mathcal{O}((K+L)N)$. Then, the total complexity of Algorithm 2 under the probabilistic location error model is given by (33), as shown at the bottom of this page. Note that the proposed Algorithm 2 also has a polynomial time computational complexity which is suitable for practical implementation.

\emph{\textbf{Remark 5:}} By comparing (19) and (33), it can be observed that $T_1$ and $T_2$ have less effect on the complexity of Algorithm 1 and Algorithm 2, respectively. Thus, the complexity difference between Algorithm 1 and Algorithm 2 is mainly due to the iteration complexity and the per-iteration computation cost. Note that the iteration complexity of Algorithm 1 is larger than that of Algorithm 2. Besides, for the per-iteration computation cost, there is a higher order polynomial term in Algorithm 1 than Algorithm 2. Thus, it can be seen that the complexity of Algorithm 2 is lower than that of the Algorithm 1 since the LMIs of size 2 in the first part of (26c) and (28c) are always hold and can be omitted in Algorithm 2.

\section{Simulation Results}
In this section, simulation results are presented to verify the performance of the proposed robust trajectory and transmit power design schemes in Section III (denoted as the bounded location scheme) and Section IV (denoted as the probabilistic location scheme), as compared to the following three benchmark schemes: 1) non-robust joint trajectory and transmit power scheme; 2) fixed trajectory scheme I; 3) fixed trajectory scheme II. Specifically, the non-robust design can be obtained from the WCR-ASRM design in Section III, where we treat the estimated location of each Eve ${{\mathbf{\bar{e}}}_{k}}$ as the actual location of Eve, e.g., $\xi_{k}=0$, and then evaluate the worst-case average secrecy rate obtained in the bounded location error model. Fixed trajectory scheme I and fixed trajectory scheme II perform robust transmit power allocation based on the bounded location error model and the probabilistic location error model, respectively, and design the UAV's trajectory in the following manner: the UAV flies straightly from the initial location to the location right above the SU with the maximum speed $V_{\max}$, then hovers there for a certain duration, and finally flies straightly to the final location with the maximum speed $V_{\max}$ by the end of time $T$. This setting is also used to generate the initial feasible points for the proposed robust schemes. According to the parameters adopted in \cite{X. Lin}, \cite{M. Cui}, we set the constant flight altitude of the UAV as $H=100$ $\text{m}$, the horizontal coordinates of the SU as $(x_s,y_s)=(0,0)$ $\text{m}$. It is assumed that there exists one PU, e.g., $L=1$, and the estimated horizontal coordinates of the PU is $(x_{p,1},y_{p,1})=(-40,-80)$ $\text{m}$. In this case,  it is possible to better observe the trend of the UAV's transmit power affected by the PU. It is assumed that there are $K=2$ Eves and the estimated horizontal coordinates of Eves are $(x_{e,1},y_{e,1})=(240,-120)$ $\text{m}$ and $(x_{e,2},y_{e,2})=(-240,120)$ $\text{m}$, respectively. Other simulation parameters are set as follows: $\rho=0.2$, $\phi=0.2$, $\varepsilon_1=5$, $\varepsilon_2=35$, $\varpi_1=5$, $V_{\max}=10$ $\text{m/s}$, $T/N=1$ $\text{s}$, $P_{\max}=4\bar{P}$, and $\sigma^2=\sigma^2_1=\sigma^2_2=-50$ $\text{dBm}$.

\begin{figure}[!t]
\centering
\includegraphics[width=3.5 in]{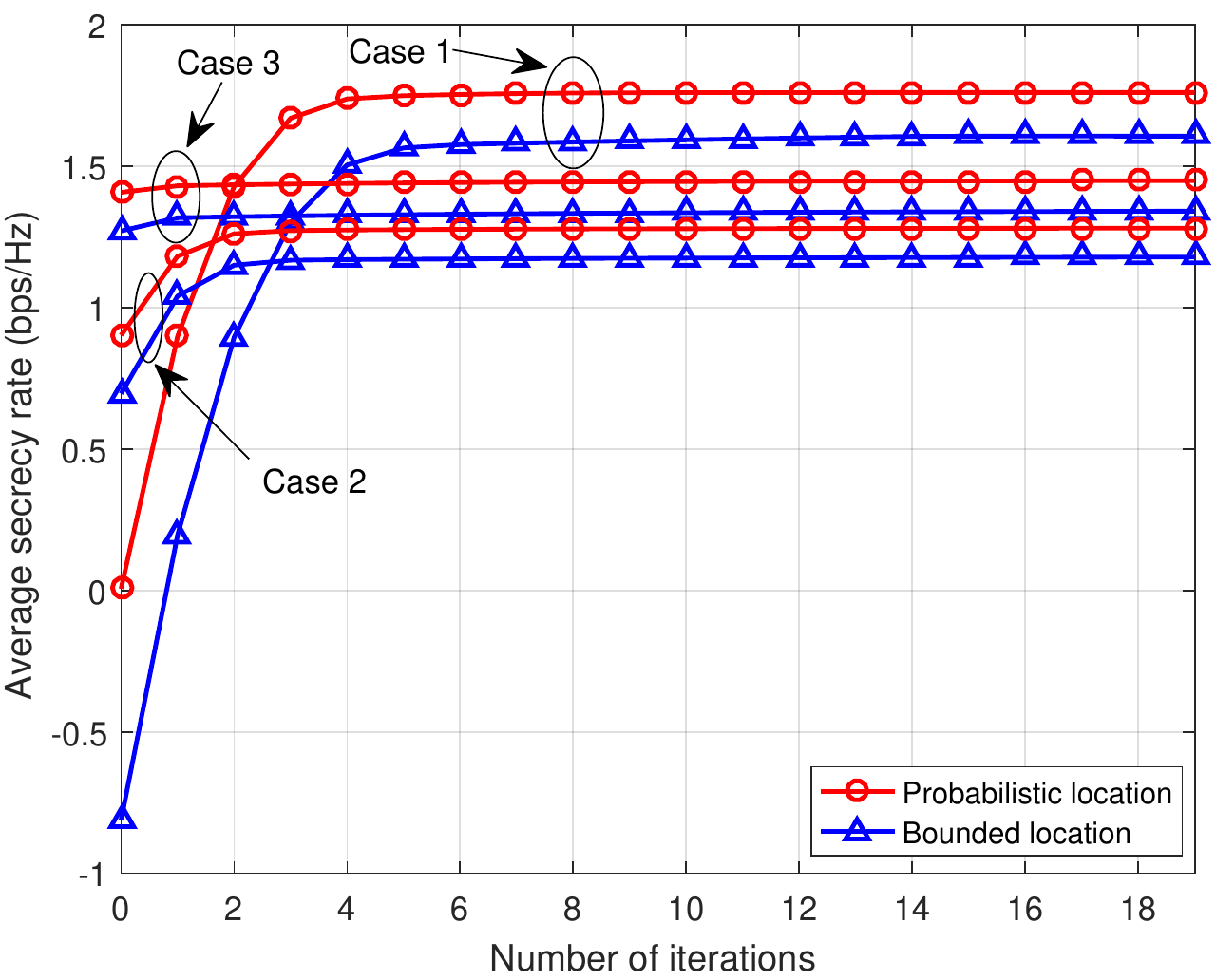}
\caption{The average secrecy rate versus the number of iterations.}
\label{fig.1}
\end{figure}

Fig. 2 shows the convergence performance of our proposed algorithms for the bounded location scheme and the probabilistic location scheme. To ensure that the UAV has enough time to fly freely and hover at desired locations, the flight time $T$ is set as $60$ s. In order to study the effect of the transmit power and the IT threshold on the performance of the UAV-enabled CR network, we consider three different cases: 1) $\bar{P}=-10$ dB, ${\Upsilon}=2.5\times10^{-7}$ W; 2) $\bar{P}=-20$ dB, ${\Upsilon}=3\times10^{-8}$ W; 3) $\bar{P}=-20$ dB, ${\Upsilon}=2.5\times10^{-7}$ W. It can been seen from Fig. 2 that the achievable average secrecy rate of case 1 is higher than that of case 3, as a higher transmit power can be used for information transfer to improve the security communication performance of the UAV system. Moreover, the superiority of the achievable average secrecy rate of case 3 over case 2 can be explained by the fact that the higher the IT threshold, the smaller the limitation of the UAV's transmit power. When the tolerable interference power threshold of the PUs is large enough, the average interference power constraint is inactive. In this case, the average interference power of the PU has no effect on the trajectory design and the power allocation. In contrast, the transmit power constraint is active and has a significant influence on the trajectory design and power allocation. When the tolerable interference power threshold of the PUs is small enough, the transmit power constraint is inactive. In this case, the average interference power constraint is active and has an effect on the trajectory design and power allocation. Thus, the UAV can achieve a higher average secrecy rate through a more flexible power allocation optimization. It is observed that the two proposed iterative algorithms can converge within a few number of iterations. It can be seen that more iterations are needed to achieve convergence when the given average transmit power of the UAV $\bar{P}$ is sufficiently large. The reason is that when the given average transmit power is large, the initial feasible points (fixed trajectory) have a great impact on the secure communication performance, resulting in a low average secrecy rate at the beginning of iterations. Thus, more iterations are needed to converge to the desired suboptimal solution. It is also observed that case 3 can converge faster than case 2. In fact, due to the higher IT threshold in case 3, the existence of the PU does not affect the maximum average secrecy rate optimization problem, resulting in a faster convergence than case 2.

\begin{figure}[!t]
\centering
\subfigure[]{
\begin{minipage}[t]{1\linewidth}
\centering
\includegraphics[width=3.5 in]{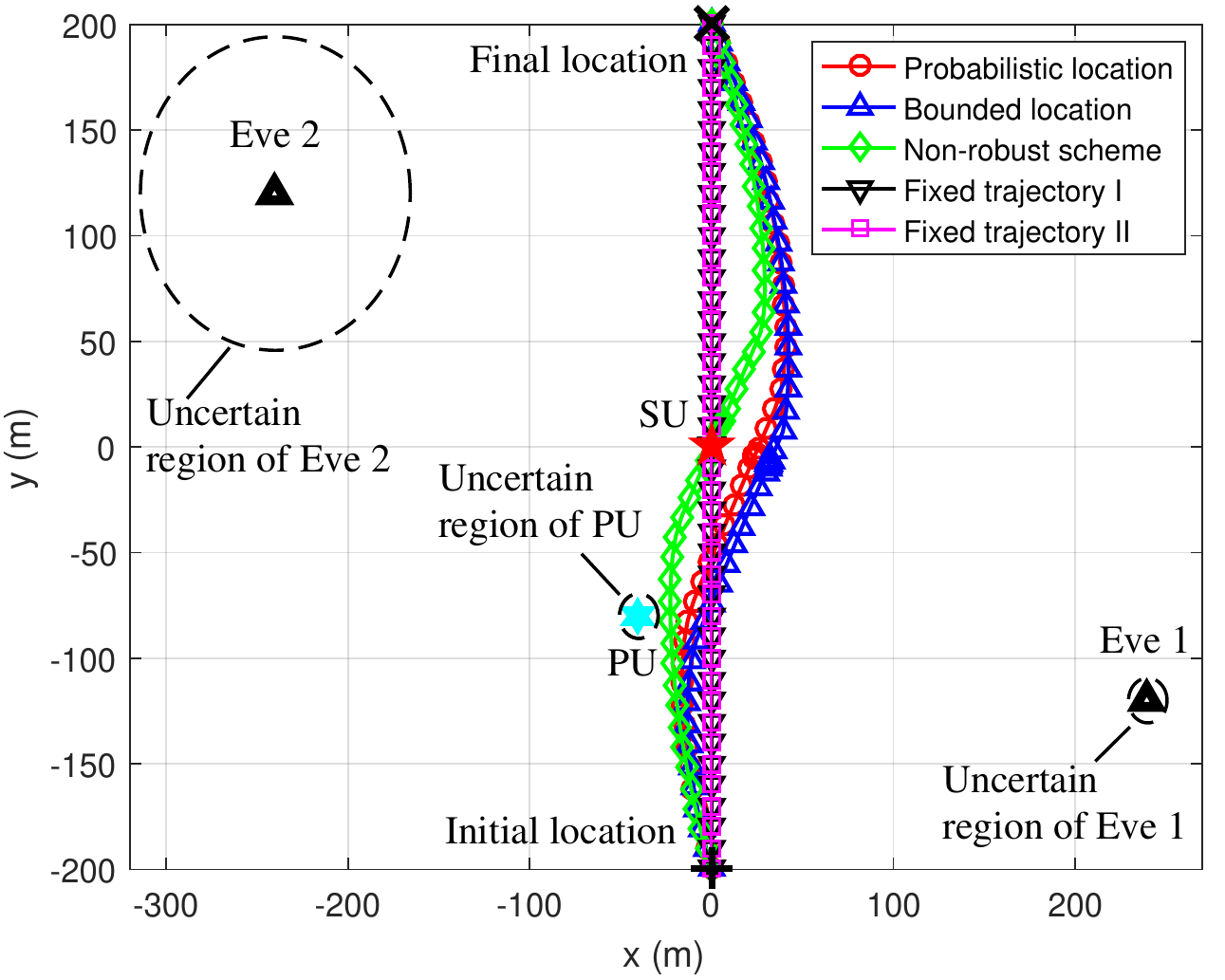}
\end{minipage}
}
\subfigure[]{
\begin{minipage}[t]{1\linewidth}
\centering
\includegraphics[width=3.5 in]{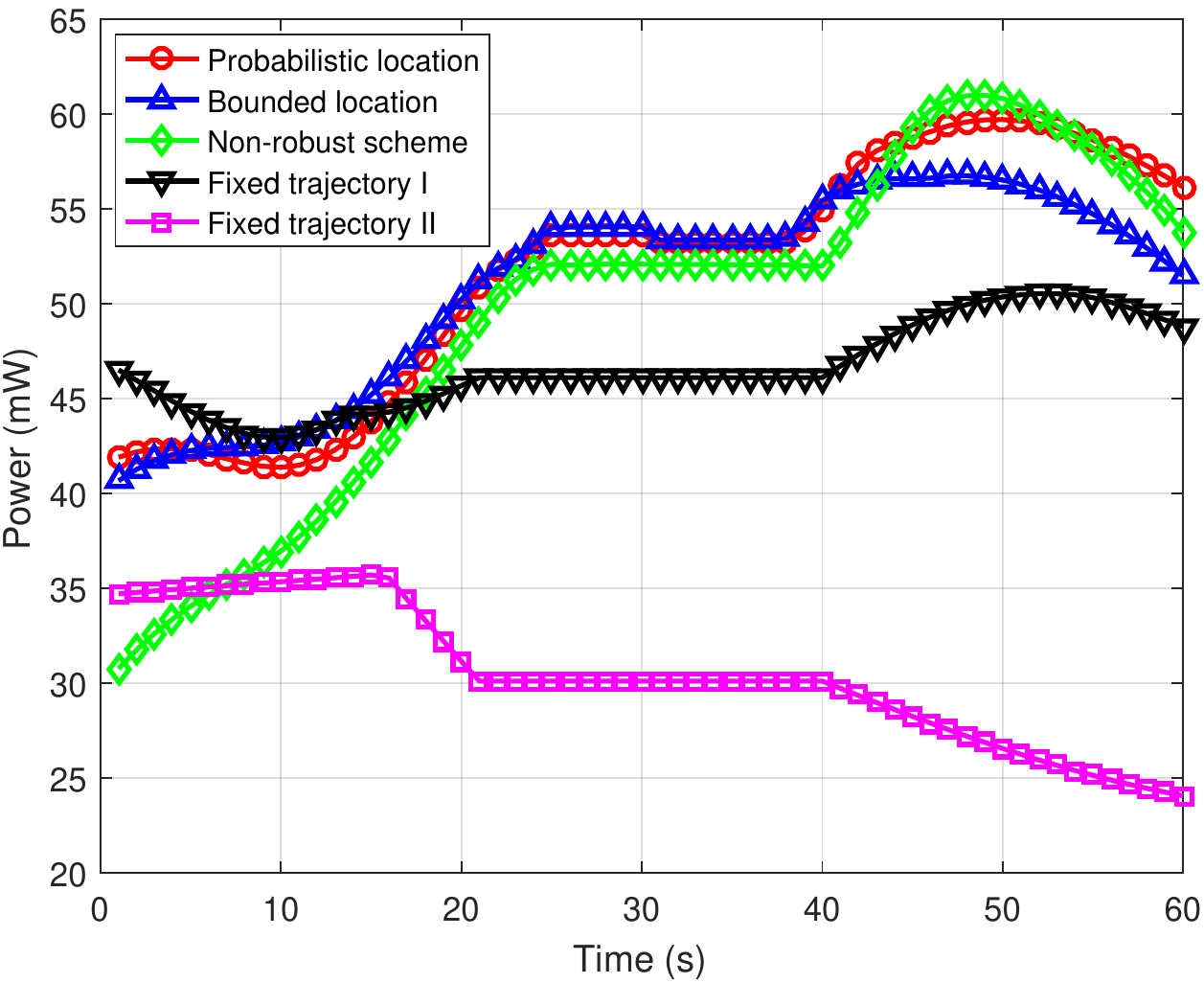}
\end{minipage}
}
\caption{(a) The UAV's trajectories for different schemes; (b) The UAV's transmit power versus the flight time.}
\end{figure}

Fig. 3(a) shows the UAV's trajectories obtained by different schemes when $T=60$ s, $\bar{P}=-10$ dB, and ${\Upsilon}=2.5\times10^{-7}$ W. Fig. 3(b) shows the UAV's transmit power versus the flight time $T$ under different schemes. It can be seen from Fig. 3(a) that the UAV's trajectory obtained by the proposed probabilistic location scheme is similar to that obtained by the bounded location scheme, and the UAV first flies towards the SU along an arc path away from Eve 1 at a high speed, then hovers over a point near the SU for a period of time, and finally flies along an arc path away from Eve 2 to the final location for the rest of the time. Note that a hovering point is the point where the maximum secrecy rate can be achieved among all the points contributing the UAV's trajectory. However, for the non-robust design, the hovering point is set directly above the SU. The reason is that the non-robust scheme ignores the uncertainty of all Eves locations. Thus, the UAV cannot obtain the desirable hover point through the non-robust scheme, resulting in an inefficient communication and a severe performance loss. Moreover, as can be seen from Fig. 3, the average interference power constraint of the PU has a low impact on the UAV's trajectory but has a significant impact on the UAV's transmit power. This shows that the trajectory optimization of the UAV under the line-of-sight channel plays an important role in the improvement of the secrecy rate of the SU. Thus, the UAV chooses to reduce the transmit power rather than stay away from the PU to satisfy the average interference power constraint of the PU. In Fig. 3(b), the UAV's transmit power is relatively small in the first 20 seconds. The reason is that in the first 20 seconds of the flight, the UAV is close to the PU, and the UAV reduces its transmit power to reduce the interference caused to the PU. In contrast, the UAV's transmit power obtained by fixed trajectory scheme II is relatively small in the last 20 seconds. Comparing to fixed trajectory scheme I, it can be seen that fixed trajectory scheme II only optimizing the transmit power is more difficult to satisfy the corresponding outage constraint under the probabilistic location error model. Hence, due to the large uncertainty of Eve 2, the UAV under fixed trajectory scheme II should significantly reduce the transmit power when approaching Eve 2. Furthermore, the actual average transmit power of the UAV is less than the given transmit power limit $\bar{P}$ due to the existence of the average interference power constraints. Also, it can be seen that the average transmit power of the proposed schemes is larger than that of the benchmark schemes. The reason is that the proposed probabilistic location scheme and the bounded location scheme both have a stronger ability to utilize energy resources than those three benchmark schemes while satisfying the QoS requirement of the PU.

Fig. 4 shows the average secrecy rate versus the total flight time under different schemes when $\bar{P}=-10$ dB and ${\Upsilon}=2.5\times10^{-7}$ W. It is seen that the average secrecy rate increases with the total flight time, regardless of the schemes. The reason is that as the total flight time increases, the UAV has more time to hover over the desirable point to achieve a more efficient communication which improves the average secrecy rate. It is also observed that as the total flight time increases, the gaps of the average secrecy rate between the proposed schemes and the benchmark schemes are enlarged. In fact, a longer flight time magnifies the performance loss since the UAV in benchmark schemes hovers on the undesired location for communication. In addition, it can be seen that the achievable average secrecy rate of the probabilistic location scheme is higher than that of the bounded location scheme. It can be explained by the fact that the secure performance of the UAV-enabled CR system achieved by the bounded location scheme is too conservative due to the worst-case objective function and stringent constraints.

\begin{figure}[!t]
\centering
\includegraphics[width=3.5 in]{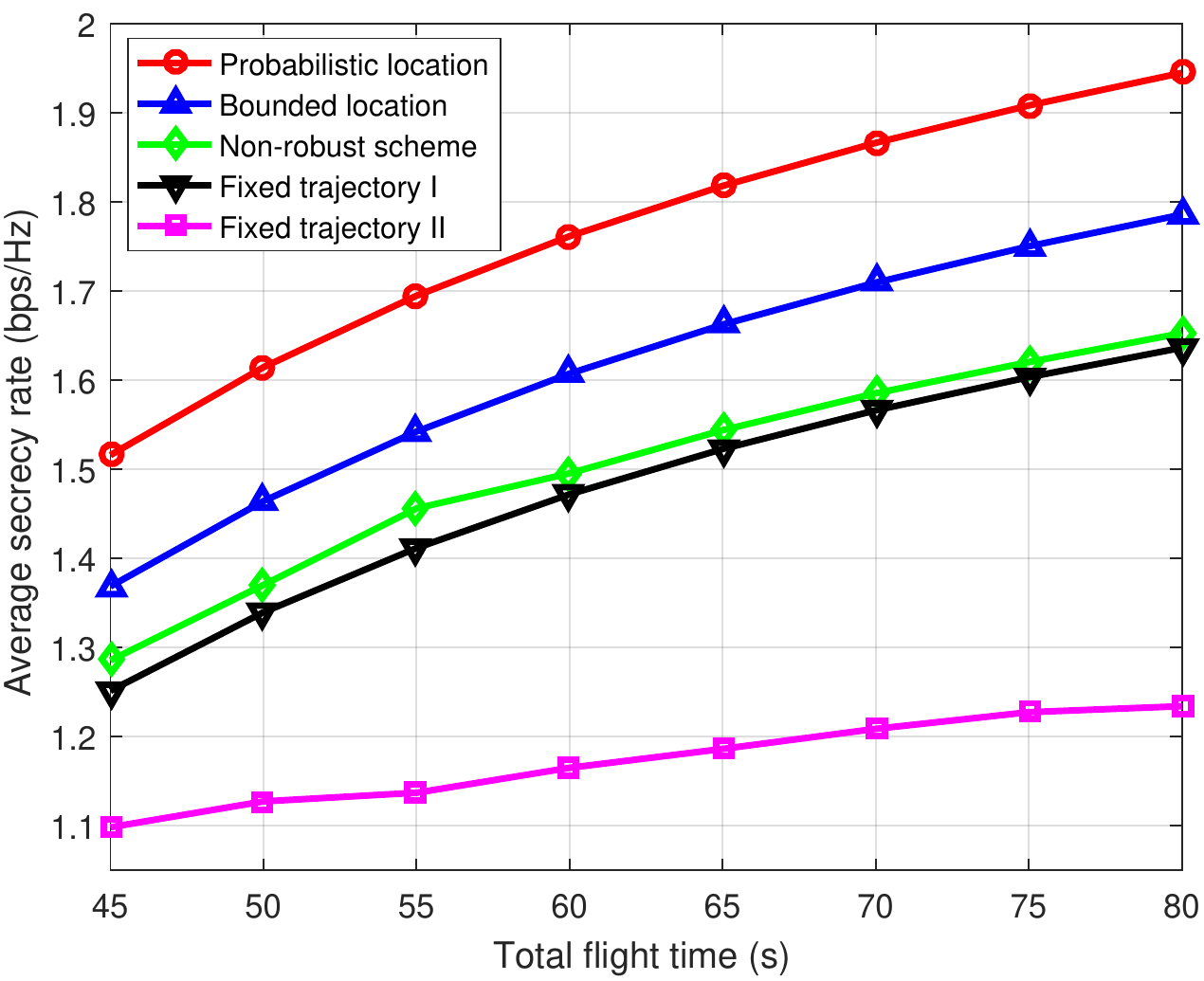}
\caption{The average secrecy rate versus the total flight time.}
\label{fig.1}
\end{figure}

\begin{figure}[!t]
\centering
\includegraphics[width=3.5 in]{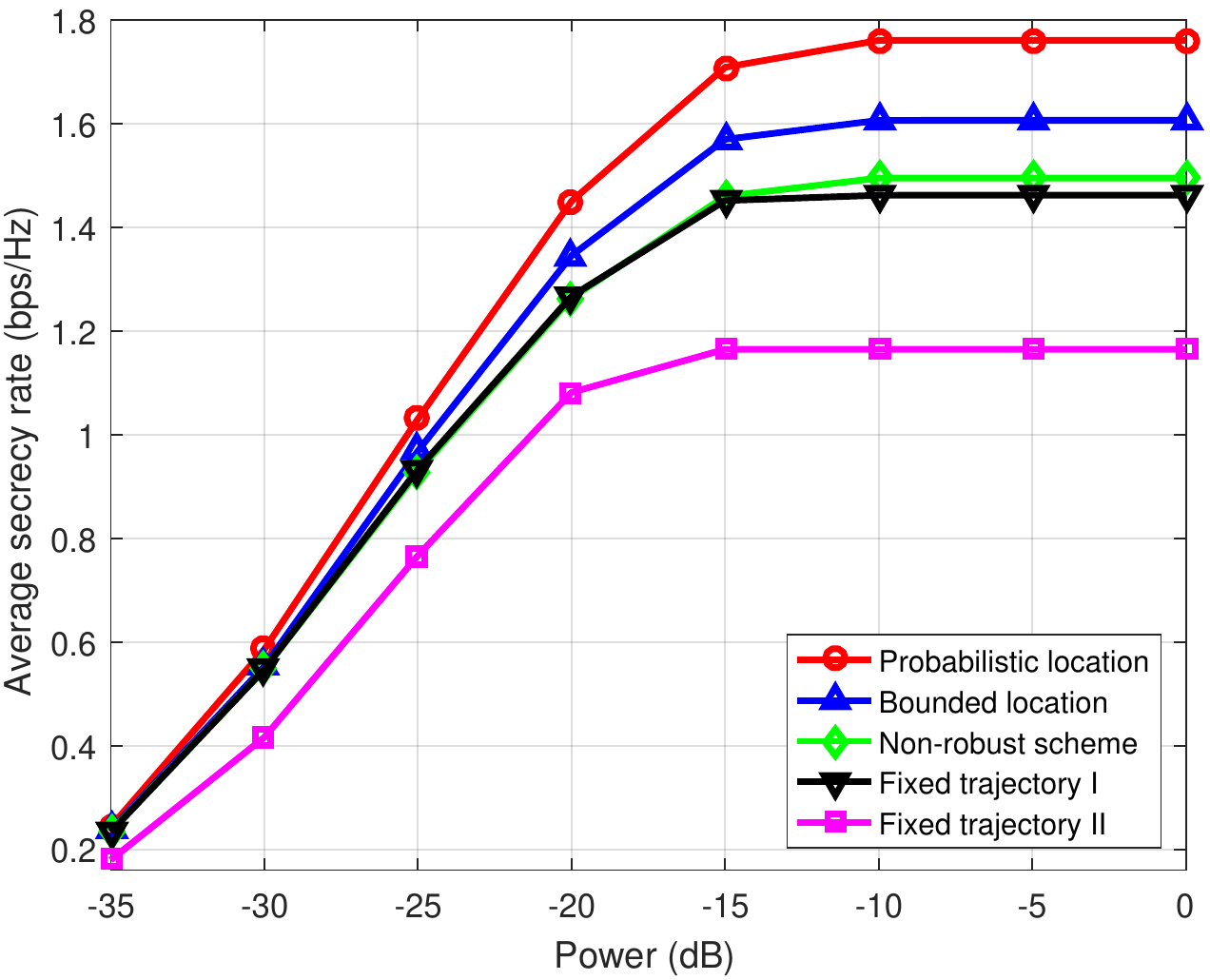}
\caption{The average secrecy rate versus the UAV's average transmit power.}
\label{fig.1}
\end{figure}

\begin{figure}[!t]
\centering
\includegraphics[width=3.5 in]{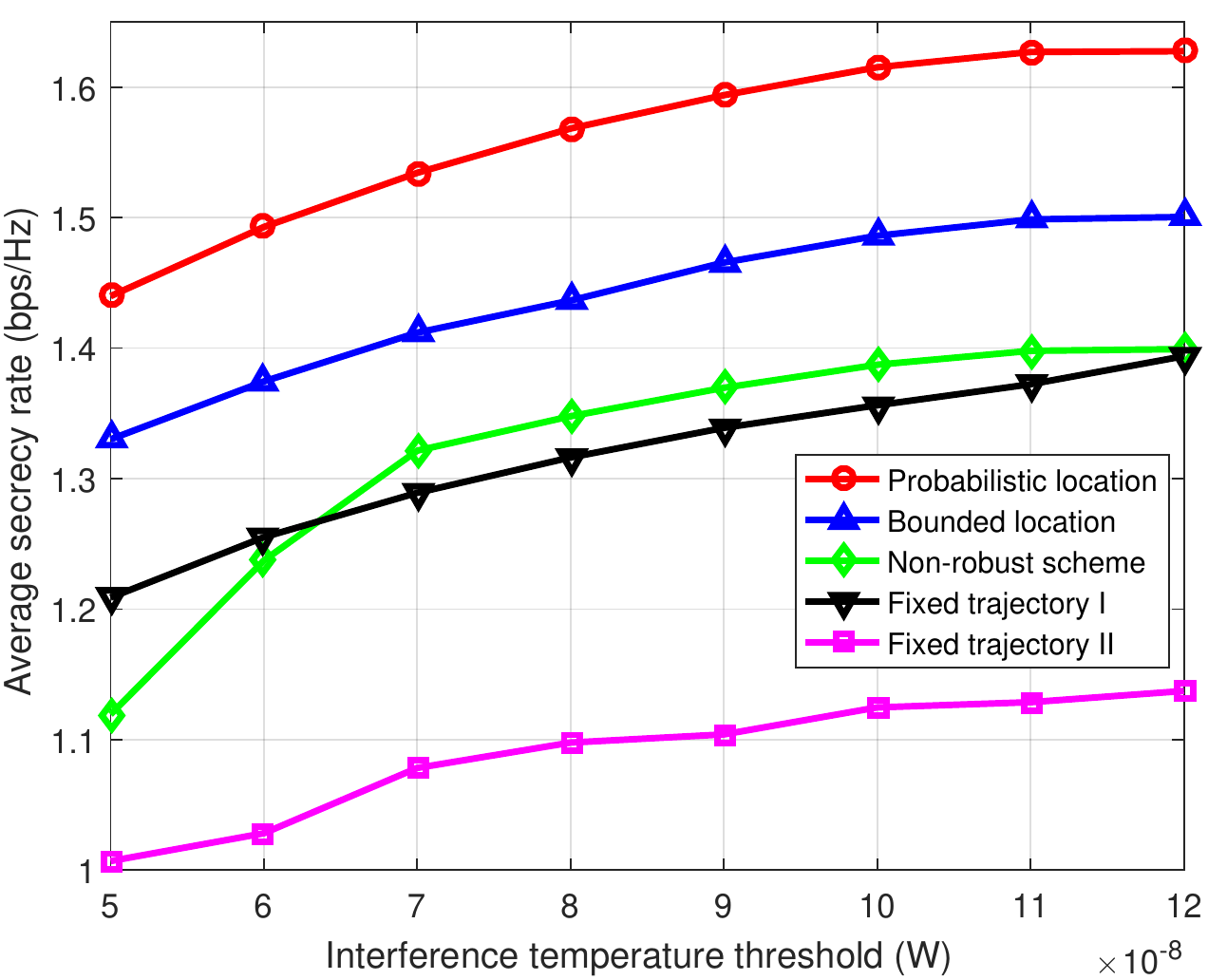}
\caption{The average secrecy rate versus the interference temperature threshold.}
\label{fig.1}
\end{figure}

Fig. 5 shows the average secrecy rate versus the UAV's average transmit power under different schemes when $T=60$ s and ${\Upsilon}=2.5\times10^{-7}$ W. It can be seen that when the UAV's average transmit power is lower than -15 dB, the average secrecy rate first increases rapidly with the transmit power. However, when the UAV's average transmit power is higher than -10 dB, the average secrecy rate becomes saturated. The reason is that with the increasing transmit power in the low transmit power regime, the UAV can fully exploit the benefit brought by the optimized trajectory to establish a efficient communication. In contrast, when $\bar{P}$ is sufficiently large, the UAV's transmit power is limited by the IT threshold ${\Upsilon}$. Thus, the average secrecy rate is mainly determined by the UAV's trajectory. Particularly, it is seen that the average secrecy rates achieved by fixed trajectory scheme I and II become saturated when the UAV's average transmit power is higher than -15 dB. This indicates that compared with other schemes, fixed trajectory schemes I and II result in a larger interference to the PU and obtain a lower energy efficiency. It is also observed that the gap of the achievable average secrecy rate between the bounded location scheme and the non-robust scheme becomes larger as the average transmit power increases. In fact, the high transmit power intensifies the performance losses as the non-robust scheme ignores the location estimation errors of Eves. Thus, as the average transmit power increases, the deviation of the non-robust optimized trajectory and the proposed robust optimized trajectory becomes larger, resulting in a large performance loss.

Fig. 6 shows the average secrecy rate versus the interference temperature threshold under different schemes when $T=60$ s and $\bar{P}=-17$ dB. It can be seen that the average secrecy rate increases with the IT threshold since more transmit power can be utilized to improve the communication performance. Note that when the IT threshold is larger than $11\times10^{-8}$ W, the average secrecy rates achieved by all schemes become saturated except for fixed trajectory schemes I and II. This indicates that fixed trajectory schemes cannot fully utilize the given average transmit power at ${\Upsilon}=11\times10^{-8}$ W. It is also seen that the achievable average secrecy rate of the non-robust scheme is rapidly reduced when the IT threshold is less than $7\times10^{-8}$ W. The reason is that when the noise power that the PU can tolerate is very low, the power allocation of the UAV is severely limited. In this case, the non-robust scheme ignores the uncertainty of all Eves locations, which may result in undesirable UAV's trajectories and inefficient communication links, further leading to a more serious degradation of communication performance.

\section{Conclusion}
Secure communication was studied in a UAV-enabled CR network. Robust trajectory and transmit power were jointly designed to realize secure communication while protecting the PU from the harmful interference under the bounded location error model and the probabilistic location error model. Two iterative algorithms based on these two location error models were proposed to obtain a suboptimal solution of the formulated non-convex problems. Theoretical derivations and simulation results showed that the proposed probability location scheme not only has lower algorithm complexity, but also achieves a higher average secrecy rate compare to the proposed bounded location scheme. Moreover, it was shown that the average secrecy rate obtained by our proposed robust trajectory and transmit power schemes are higher than those achieved by the benchmark schemes.
\appendices
\section{Proof of Proposition 1}
The Hessian matrix of $f\left( x,y \right)$, where $x>0$ and $y>0$, is given as
\setcounter{equation}{33}
\begin{align} \notag
&{{\nabla }^{2}}f\left( x,y \right)=\frac{1}{xy}\left[ \begin{matrix}
   2{{x}^{-2}} & {{\left( xy \right)}^{-1}}  \\
   {{\left( xy \right)}^{-1}} & 2{{y}^{-2}}  \\
\end{matrix} \right]\\
&=\frac{1}{xy}\left[ \begin{matrix}
   {{x}^{-2}} & 0  \\
   0 & {{y}^{-2}}  \\
\end{matrix} \right]+\frac{1}{xy}\left[ \begin{matrix}
   {{x}^{-1}}  \\
   {{y}^{-1}}  \\
\end{matrix} \right]\left[ \begin{matrix}
   {{x}^{-1}} & {{y}^{-1}}  \\
\end{matrix} \right]\succeq\mathbf{0}.
\end{align}
Note that the Hessian matrix of $f\left( x,y \right)$ can be rewritten as a linear combination of two positive semidefinite matrices. Thus, $f\left( x,y \right)$ is a convex function. This completes the proof.

\begin{IEEEbiography}[{\includegraphics[width=1.0in,height=1.15in,clip,keepaspectratio]{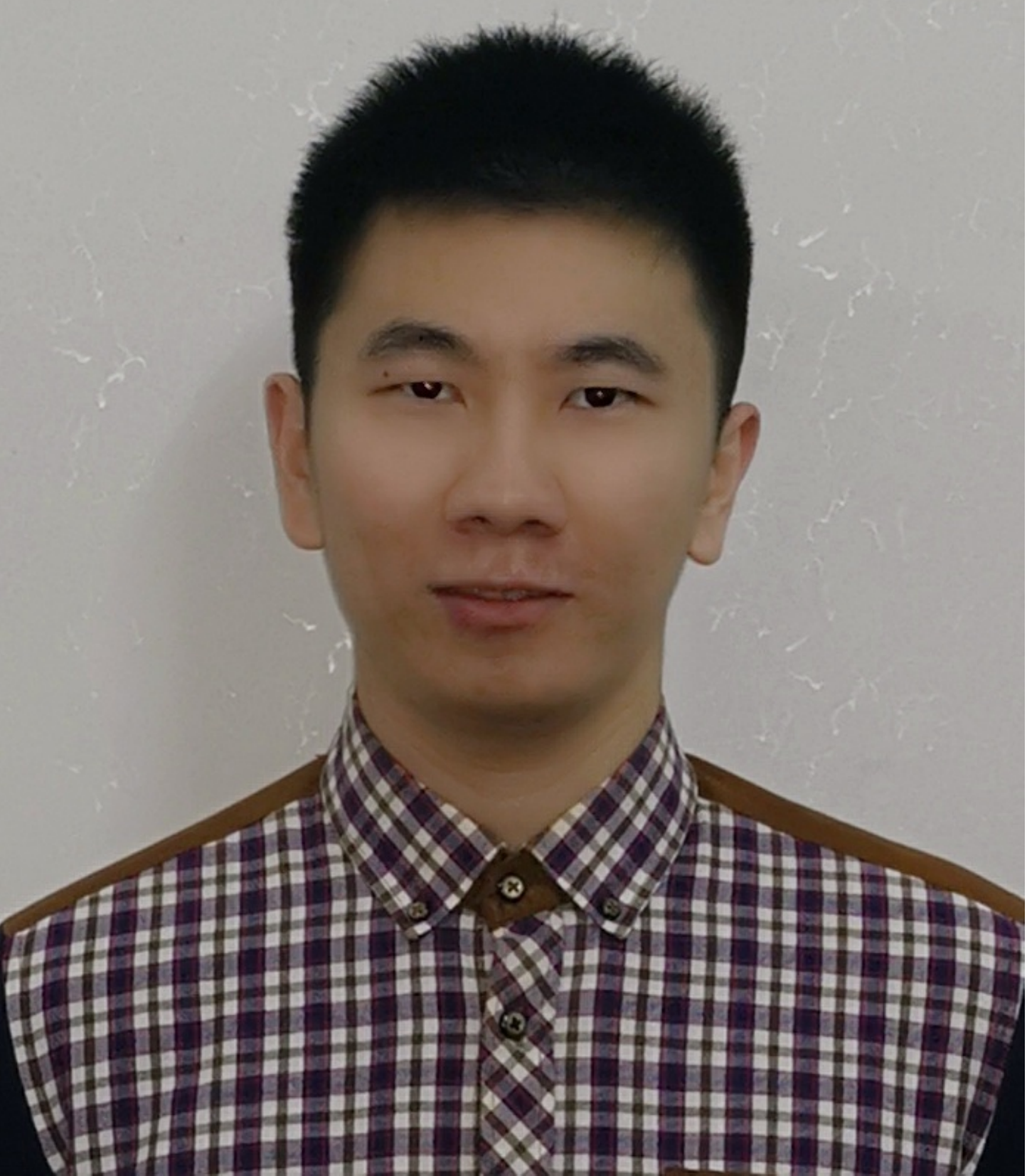}}]{Yifan Zhou}
received the bachelor's degree from Jiangxi University of Finance and Economics, Jiangxi, China, in 2017. He is currently pursuing the master's degree with the School of Information Engineering, Nanchang University, Jiangxi, China. His research interests focus on 5G wireless networks, cognitive radio, physical layer security, UAV-enabled communication.
\end{IEEEbiography}

\begin{IEEEbiography}[{\includegraphics[width=1.0in,height=1.15in,clip,keepaspectratio]{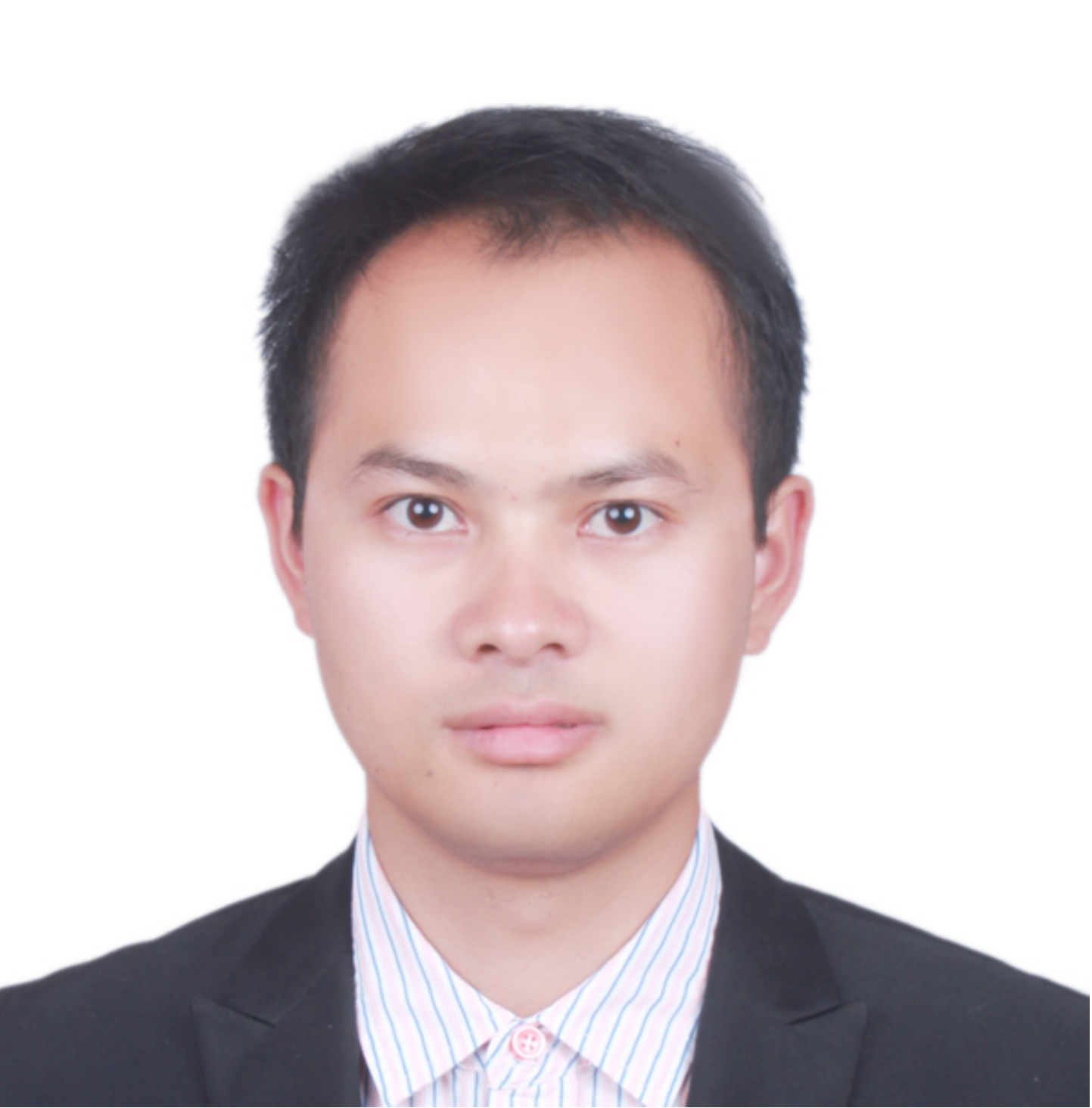}}]{Fuhui Zhou}
has worked as a Senior Research Fellow at Utah State University. He received the Ph. D. degree from Xidian University, Xian, China, in 2016. He is currently a Full Professor at College of Electronic and Information Engineering, Nanjing University of Aeronautics and Astronautics. His research interests focus on cognitive radio, edge computing, machine learning, NOMA, physical layer security, and resource allocation. He has published more than 90 papers, including IEEE Journal of Selected Areas in Communications, IEEE Transactions on Communications, IEEE Wireless Communications, IEEE Network, IEEE GLOBECOM, etc. He was awarded as Young Elite Scientist Award of China. He has served as Technical Program Committee (TPC) member for many International conferences, such as IEEE GLOBECOM, IEEE ICC, etc. He serves as an Editor of IEEE Transactions on Communications and IEEE Wireless Communications Letters, and an Associate Editor of IEEE Systems Journal, IEEE Access and Physical Communications. He also serves as co-chair of IEEE Globecom 2019 and IEEE ICC 2019 workshop on Advanced Mobile Edge /Fog Computing for 5G Mobile Networks and Beyond.
\end{IEEEbiography}

\begin{IEEEbiography}[{\includegraphics[width=1.0in,height=1.28in,clip,keepaspectratio]{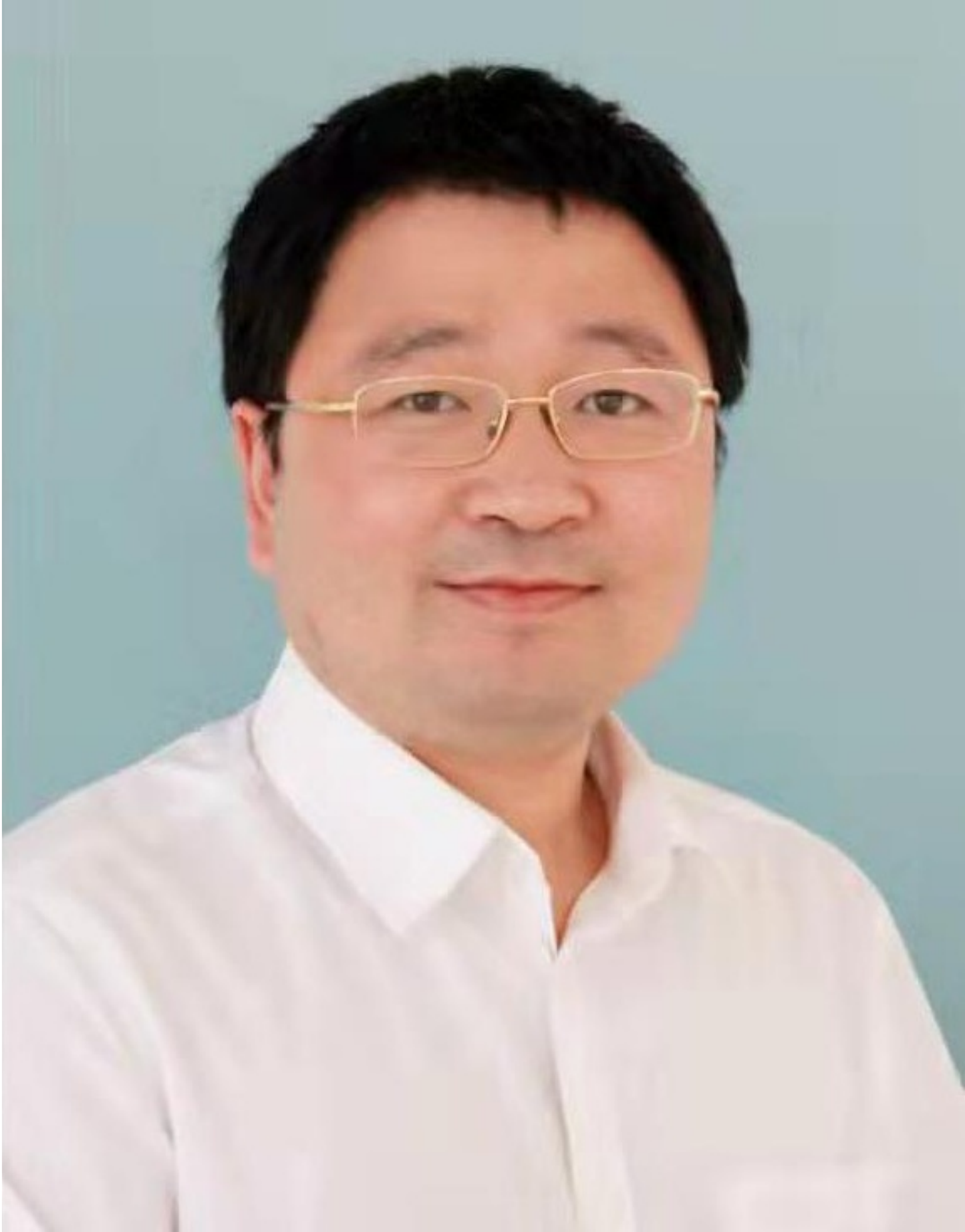}}]{Huilin Zhou}
was born in China in 1979. He received the Ph.D. degree in information engineering from Wuhan University, Wuhan, China.

He is currently a Professor with School of Information Engineering, Nanchang University, Nanchang, China. His research interests include radar system, radar signal processing, and radar imaging.
\end{IEEEbiography}

\begin{IEEEbiography}[{\includegraphics[width=0.95in,height=1.4in]{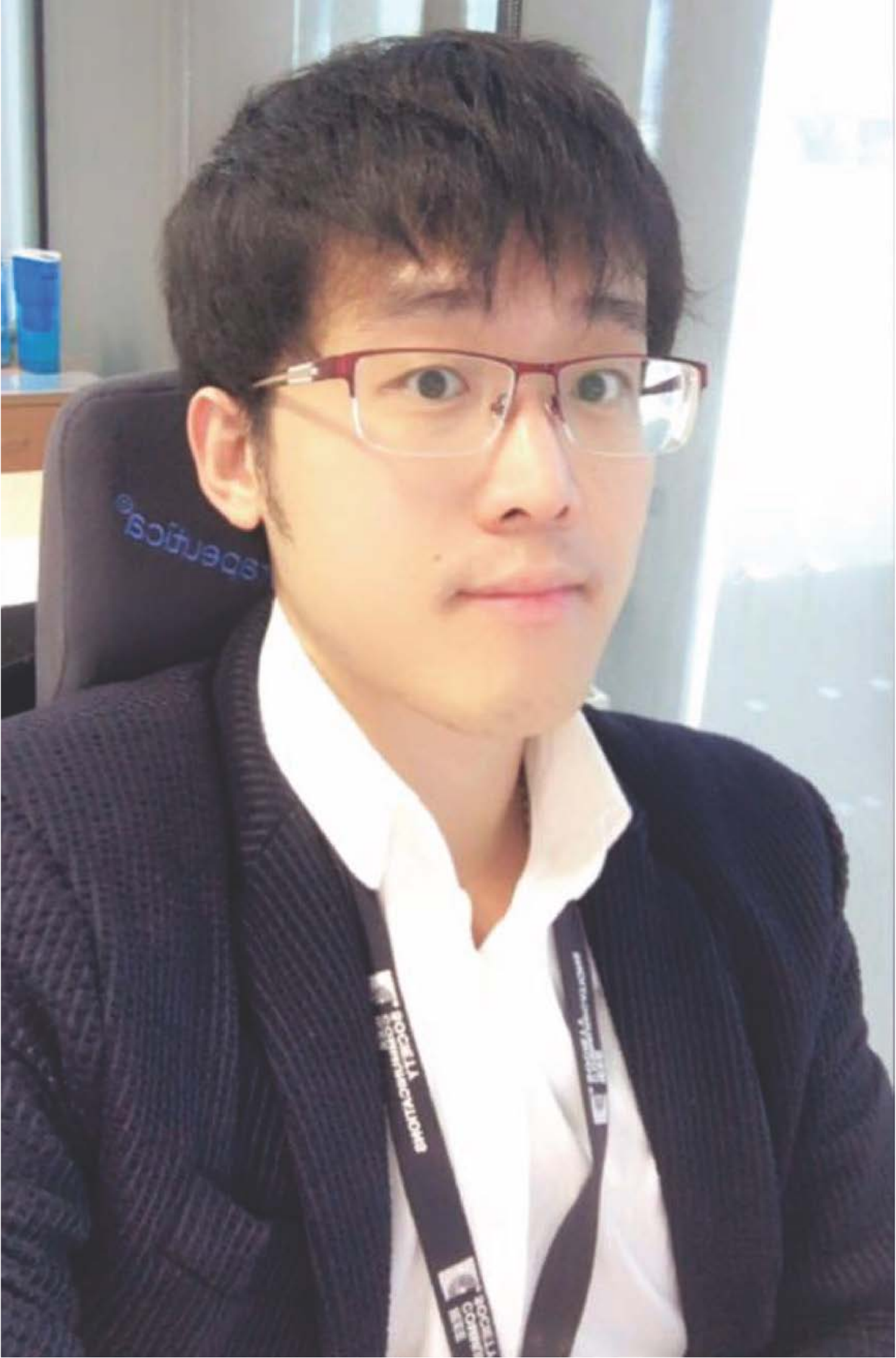}}]{Derrick
Wing Kwan Ng (S'06-M'12-SM'17)  } received the bachelor degree with first-class honors and the Master of Philosophy (M.Phil.) degree in electronic engineering from the Hong Kong University of Science and Technology (HKUST) in 2006 and 2008, respectively. He received his Ph.D. degree from the University of British Columbia (UBC) in 2012. He was a senior postdoctoral fellow at the Institute for Digital Communications, Friedrich-Alexander-University Erlangen-N\"urnberg (FAU), Germany. He is now working as a Senior Lecturer and a Scientia Fellow at the University of New South Wales, Sydney, Australia.  His research interests include convex and non-convex optimization, physical layer security, IRS-assisted communication, UAV-assisted communication, wireless information and power transfer, and green (energy-efficient) wireless communications.
Dr. Ng received the Best Paper Awards at the IEEE TCGCC Best Journal Paper Award 2018, INISCOM 2018, IEEE International Conference on Communications (ICC) 2018,  IEEE International Conference on Computing, Networking and Communications (ICNC) 2016,  IEEE Wireless Communications and Networking Conference (WCNC) 2012, the IEEE Global Telecommunication Conference (Globecom) 2011, and the IEEE Third International Conference on Communications and Networking in China 2008.  He has been serving as an editorial assistant to the Editor-in-Chief of the IEEE Transactions on Communications from Jan. 2012 to Dec. 2019. He is now serving as an editor for the IEEE Transactions on Communications,  the IEEE Transactions on Wireless Communications, and an area editor for the IEEE Open Journal of the Communications Society. In addition, he is listed as a Highly Cited Researcher by Clarivate Analytics in 2018 and 2019.
\end{IEEEbiography}

\begin{IEEEbiography}[{\includegraphics[width=1.0in,height=1.15in,clip,keepaspectratio]{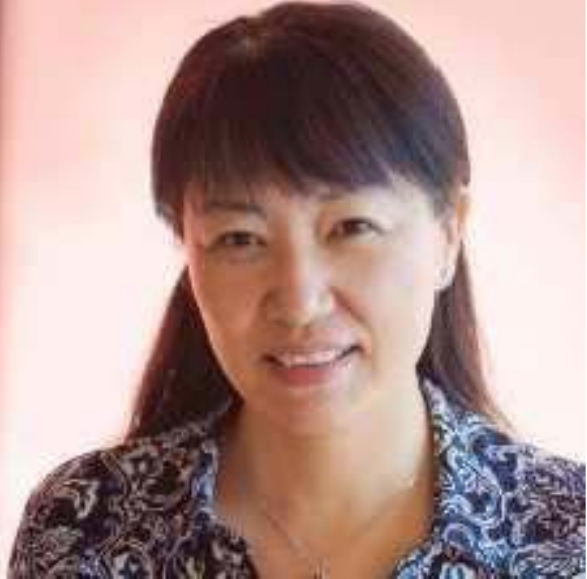}}]{Rose Qingyang Hu}
 (S'95-M'98-SM'06-F'19) is a Professor in the Electrical and Computer Engineering Department and Associate Dean for research of College of Engineering  at Utah State University. She also directs Communications Network Innovation Lab at Utah State University. Her current research interests include next-generation wireless system design and optimization, Internet of Things, Cyber Physical system, Mobile Edge Computing, V2X communications, artificial intelligence in wireless networks, wireless system modeling and performance analysis. Prof. Hu received the B.S. degree from the University of Science and Technology of China, the M.S. degree from New York University, and the Ph.D. degree from the University of Kansas. Besides a decade academia experience, she has more than 10 years of R\&D experience with Nortel, Blackberry, and Intel as a Technical Manager, a Senior Wireless System Architect, and a Senior Research Scientist, actively participating in industrial 3G/4G technology development, standardization, system level simulation, and performance evaluation. She has published extensively and holds numerous patents in her research areas. Prof. Hu is currently serving on the editorial boards of the IEEE Transactions on Wireless Communications, the IEEE Transactions on Vehicular Technology, the IEEE Communications Magazine and the IEEE Wireless Communications.  She also served as the TPC Co-Chair for the IEEE ICC 2018. She is an IEEE Communications Society Distinguished Lecturer Class 2015-2018.  She was a recipient of prestigious Best Paper Awards from the IEEE GLOBECOM 2012, the IEEE ICC 2015, the IEEE VTC Spring 2016, and the IEEE ICC 2016. Prof. Hu is senior member of IEEE and a member of Phi Kappa Phi Honor Society.
\end{IEEEbiography}

\end{document}